\documentclass[sigconf]{acmart}

\usepackage{graphicx}
\usepackage{subfig}
\usepackage{amsmath}
\usepackage{nicefrac}
\usepackage{listings}
\usepackage{appendix}
\usepackage{xspace}
\usepackage{xfrac}
\usepackage{etoolbox}
\usepackage{booktabs} % For formal tables
\usepackage[utf8]{inputenc}
\usepackage{xcolor}
\usepackage{verbatim}
\PassOptionsToPackage{hyphens}{url}
\usepackage{hyperref}
\usepackage{multirow}
\usepackage{colortbl}
\usepackage{microtype}
\usepackage{balance}
\usepackage{algorithm,algorithmicx,algpseudocode}
\usepackage{caption}

\newcommand{\TheSystem}{Cortex\xspace}
\newcommand{\NewPara}[1]{\vspace{4pt}\noindent{\bf #1}}
\newcommand{\Section}[1]{\S\ref{sec:#1}}
\newcommand{\Equation}[1]{Eq.~\ref{eq:#1}}
\newcommand{\Figure}[1]{Fig.~\ref{fig:#1}}
\newcommand{\Table}[1]{Tab.~\ref{tab:#1}}

\newcommand{\ma}[1]{\textnormal{\color{red}{\textbf{Mohammad: }\textbf{#1}}}\unskip}

\def\compactify{\itemsep=0pt \topsep=0pt \partopsep=0pt \parsep=0pt \leftmargin=0.5cm}
\let\latexusecounter=\usecounter

\newenvironment{CompactEnumerate} 
  {\def\usecounter{\compactify\latexusecounter} 
   \begin{enumerate}}
  {\end{enumerate}\let\usecounter=\latexusecounter}

\renewcommand\footnotetextcopyrightpermission[1]{}
\begin{document}
\pagestyle{plain}

\title{\TheSystem: Harnessing Correlations to Boost Query Performance}

\author{Vikram Nathan, Jialin Ding, Tim Kraska, Mohammad Alizadeh}
\affiliation{Massachusetts Insititute of Technology
}
\email{{vikramn,jialind,kraska,alizadeh}@mit.edu}



\begin{abstract}
\if 0
Databases typically employ indexes, such as clustered B-Trees, to filter out irrelevant records, which reduces scan overhead and speeds up query execution. However, this optimization is only available to queries that filter on the indexed attribute. To extend these speedups to queries on other attributes, database systems have turned to secondary indexes and, increasingly, multi-dimensional indexes. However, these approaches are restrictive: secondary indexes have a large memory footprint and can only speed up queries that access a small number of records; multi-dimensional indexes cannot scale to more than a handful of columns.
We present \TheSystem, an approach that takes advantage of correlations to extend the reach of primary indexes to index more attributes. Unlike prior work, which is extremely limited in the types of correlations it handles, \TheSystem can adapt itself to any existing primary index, whether single or multi-dimensional, to harness a broad variety of correlations, such as those that exist between more than two attributes or have a large number of outliers. We demonstrate that on real datasets exhibiting these diverse types of correlations, \TheSystem matches or outperforms traditional secondary indexes using more the $5\times$ less space. At the same time, it is $2-8\times$ faster than existing approaches to index correlations.
\fi 

Databases employ indexes to filter out irrelevant records, which reduces scan overhead and speeds up query execution. However, this optimization is only available to queries that filter on the indexed attribute. To extend these speedups to queries on other attributes, database systems have turned to secondary and multi-dimensional indexes. Unfortunately, these approaches are restrictive: secondary indexes have a large memory footprint and can only speed up queries that access a small number of records, and multi-dimensional indexes cannot scale to more than a handful of columns.
We present \TheSystem, an approach that takes advantage of correlations to extend the reach of primary indexes to more attributes. Unlike prior work, \TheSystem can adapt itself to any existing primary index, whether single or multi-dimensional, to harness a broad variety of correlations, such as those that exist between more than two attributes or have a large number of outliers. We demonstrate that on real datasets exhibiting these diverse types of correlations, \TheSystem matches or outperforms traditional secondary indexes with $5\times$ less space, and it is $2-8\times$ faster than existing approaches to indexing correlations.

\end{abstract}

\maketitle

\section{Introduction}
\label{sec:intro}

Data storage and access optimizations are at the core of any database system. Their main purpose is ensuring that records are efficiently looked up and filtered to maximize query performance.
For example, many databases use clustered B-Tree indexes on a single attribute or -- in the case of column stores -- sort the data on a single column. 
As long as queries filter on the single clustered attribute, both techniques allow the database to return all relevant records without scanning the entire table.

For queries that do not filter on the clustered column, secondary indexes can help boost performance, but at a steep cost: they have a notoriously large memory footprint and incur high latencies due to their lookup cost and non-sequential access patterns~\cite{main-mem-secondary, secondary-access-path}. This makes secondary indexes worth the high storage cost only if the queries touch a handful of records (i.e., have \textit{low selectivity}). 

% % TODO: don't make this sound like each query filters over many attributes.
% % The workload as a whole filters over many
% Database systems often encounter query workloads that filter over a variety of attributes, especially in the context of analytics. A traditional relational database system (RDMS) with only a primary clustered index would have to resort to a costly full table scan on any queries that don't filter on the clustered column.
% As a result, modern RDMSes employ secondary indexes, like B+ trees, on frequently filtered columns to accelerate those queries.
% % secondary-space: https://www.cs.cmu.edu/~dga/papers/hybrid-index-sigmod2016.pdf
% % secondary-compression: G. Graefe et al. Modern B-Tree Techniques. Foundations and Trends in Databases, 3(4):203–402, 2011.
% % x-tree: https://bib.dbvis.de/uploadedFiles/190.pdf
% % 
% Secondary indexes can help boost performance on selective queries, but at a steep cost: they incur a large space overhead, and their lookup times become prohibitive if the query accesses more than a small fraction of the table~\cite{secondary-space, secondary}. This makes secondary indexes worth the high storage cost only if the queries touch only a few records (i.e., have \textit{low selectivity}). Recent efforts to improve this tradeoff by compressing secondary indexes reduce their memory footprint slightly but slow down lookup times, hurting performance~\cite{secondary-compression}.

%TODO:cite tsunami
As a result, multi-dimensional primary indexes (and sort orders) are becoming increasingly popular, since they allow databases to improve the lookup performance for multiple attributes simultaneously without the storage or access overhead of a secondary index~\cite{amazon-zorder, oracle-zorder, spark-sql, flood,qdtree-sigmod, ibm-rtree}.
%For example,  Redshift~\cite{amazon-zorder} and SparkSQL~\cite{spark-sql} use Z-ordering to lay out the data, Microsoft uses zone-maps~\cite{MSR}, Vertica allows users to define a sort-order over multiple columns (e.g., first age, then date), and IBM Informix uses an R-Tree~\cite{ibm-rtree}.
Unfortunately, multi-dimensional indexes typically do not scale past a handful of columns; the \emph{curse of dimensionality} causes the performance of these indexes to quickly degrade as more columns are added~\cite{xtree, flood}. This limits the usefulness of multi-dimensional indexes on large tables, where query workloads may filter on many different columns.

% % etree: http://www.cs.cmu.edu/~euclid/etreeimr.pdf
% The recent shift to multi-dimensional indexes offers a different take on the same problem. Instead of creating a secondary index to handle queries on different attributes, an RDMS may choose to unify multiple columns into a more complex data structure, like a Z-order index, R-tree, or Octree~\cite{amazon-zorder, oracle-zorder, ibm-rtree, etree, spark-sql, multirestrees}. As a result, a multi-dimensional index can simultaneously index a primary key \emph{and} additional attributes, boosting performance on broad range of queries with minimal storage overhead relative to a B+ tree. Recent work demonstrates that on analytic workloads, multi-dimensional indexes can offer substantial storage and performance benefits over existing approaches, both in memory and on disk~\cite{flood,tsunami, qd-tree}.

% Unfortunately, multi-dimensional indexes typically do not scale past a handful of columns; the \emph{curse of dimensionality} causes the performance of these indexes to quickly degrade as more columns are added~\cite{xtree}. This limits the usefulness of multi-dimensional indexes on large tables, where query workloads may filter on many different columns.
% Between bulky secondary indexes and rigid multi-dimensional structures, an RDMS is left with no choice to handle these diverse workloads that is simultaneously compact, scalable, and performant.

% This work bridges the gap by harnessing the correlations that exist between attributes, which abound in real datasets. 
This work explores how \textit{correlations} between attributes can be leveraged to extend the reach of primary indexes, whether single or multi-dimensional, to more attributes.
Knowing that two columns are closely related can let us encode one in terms of the other, reducing the combined storage cost without taking a large performance hit. In tandem, this can present a compelling alternative to a secondary index by allowing databases to filter queries on a larger set of attributes without resorting to a costly full table scan.

Previous approaches to indexing correlations are very limited, both in the types of correlations they can handle, and in their ability to leverage primary indexes (especially multi-dimensional).
%Previous approaches to indexing correlations are not equipped to exploit the diverse types of correlations found in real world data.
For example, 
Correlation Maps (CMs)~\cite{cm} can only support strong correlations, and fall back to scanning a large fraction of the table in the presence of outliers.
Hermit~\cite{hermit} can only handle outliers for strong (i.e., \textit{soft-functional}) correlations between two real-valued attributes, a restrictive subset of the possible correlation types. Moreover, it only captures correlations between unclustered attributes and cannot capitalize on the efficiency of primary index scans.
These shortcomings limit the benefits of prior solutions in real-world settings, where correlations in data are complex and scans over primary indexes are often substantially faster than secondary indexes.

% However, past attempts to take advantage of correlations suffer from one or more of the following drawbacks:
% \begin{CompactEnumerate}
% \item Their performance degrades substantially in the presence of outliers, which are prevalent in real-world data.
% \item They can only index strong (i.e. \emph{soft-functional}) correlations between two real-valued attributes, a restrictive subset out of the variety of possible correlation types.
% \end{CompactEnumerate}
% These shortcomings diminish the applicability of prior solutions to real datasets, which naturally have outliers and contain varied and complex correlations.

We present \textbf{\TheSystem}, a general approach for indexing correlations that is the first to be robust to diverse types of correlations, including those littered with outliers and those between more than two attributes.
\TheSystem works in tandem with any primary index that already exists on the table, either single or multi-dimensional, and can adapt itself to the primary index to optimize its performance.
\TheSystem does not pay a price for its versatility: it demonstrates improved space utilization compared to an optimized secondary index, while achieving more competitive query times than existing solutions.

Key to \TheSystem's size and speed advantages are two core ideas. First, it is designed to operate jointly with a primary index (or sort order), leveraging it for fast range scans whenever possible while occasionally issuing point lookups to a small secondary index ``stash'' for outliers. Second, \TheSystem redefines what it means for a point to be an outlier. In contrast to previous solutions, which define outliers relative to a hard-coded (usually linear) model, \TheSystem makes \emph{no a priori assumptions} about the nature of a correlation. Instead, it defines outliers through the lens of {\em performance}: outliers are exactly those records that, when stashed, strike the best balance between minimizing unnecessary scans on the primary index and the stash size. 
This definition enables \TheSystem to automatically navigate the tradeoff between performance and storage overhead over a wide range of query selectivities and primary index structures.
This new definition is also responsible for \TheSystem's ability to handle arbitrary correlations: by linking the classification of outliers only to performance and not to manually configured models, \TheSystem does not suffer from the restrictions of previous approaches.

To summarize, this paper makes the following contributions:
\begin{CompactEnumerate}
\item We design and implement \TheSystem, a system that indexes correlated columns by leveraging the table's primary index. \TheSystem can use any type of primary index or sort order, including those that are multi-dimensional. In line with the shift to in-memory databases~\cite{cheap-ram}, \TheSystem is implemented on top of a read-optimized compression-enabled in-memory column store, and supports insertions, updates, and deletions.
\item We introduce a new definition of an outlier that caters specifically to database performance, along with an algorithm to classify points as outliers using this definition. This algorithm allows \TheSystem to handle multiple types of correlations and adapt to different primary indexes.
\item We use three real datasets to evaluate \TheSystem against an optimized secondary B-Tree index and previous correlation indexes. We show that \TheSystem is the most performant index on a wide range of selectivities. Notably, it matches a B-Tree's performance on low selectivity queries while using more than $5\times$ less space, and outperforms it on high selectivity queries by more than $3\times$. Additionally, \TheSystem outperforms both CMs and Hermit across the board by $2-8\times$.
%\item We demonstrate \TheSystem's ability to handle dynamic updates, scale to large numbers of columns, and effectively handle diverse types of correlations without manual tuning.
\end{CompactEnumerate}

\section{Background}
\label{sec:corr}
% db2-opt: http://ftp.cse.buffalo.edu/users/azhang/disc/SIGMOD/pdf-files/582/723-exploiting.pdf
% coradd: https://www.vldb.org/pvldb/vldb2010/pvldb_vol3/R98.pdf
% cords: https://cs.uwaterloo.ca/~ilyas/papers/cords.pdf
\TheSystem's goal is to use correlations to boost query performance specifically by accelerating data access.
There is a substantial body of work on discovering correlations in a dataset and using them for query plan selection or selectivity estimation~\cite{cords, coradd, db2-opt}. Both are outside the scope of this paper: while an algorithm for finding correlations would complement \TheSystem, our focus is on limiting the number of records accessed to reduce scan overhead. In this section, we examine past work on indexing correlations, why using correlations to speed up queries is difficult, and the gaps that remain between the state-of-the-art and a practical, efficient solution.

% mongo-partial: https://docs.mongodb.com/manual/core/index-partial/
% partial-index: https://dsf.berkeley.edu/papers/ERL-M89-17.pdf
% postgres-partial: https://www.postgresql.org/docs/8.0/indexes-partial.html
Suppose a table has two columns, $A$ and $B$, with an index on only $B$. In a typical RDMS, a database administrator wishing to handle queries over $A$ might use a secondary index structure, such as a B-Tree~\cite{btree}, to map values of $A$ to the location of the corresponding records. However, secondary indexes are well-known to be extremely memory intensive and inefficient for even moderately selective queries, since the random point accesses they incur are much slower than scanning a contiguous range of records~\cite{secondary-access-path}. Efforts to reduce their memory footprint generally deteriorate lookup performance~\cite{main-mem-secondary, partial-index}. Other approaches index only a subset of records, based on whether they satisfy a particular condition; however, these approaches only support simple predicates, have to be manually specified, and can only be utilized for a limited number of queries~\cite{partial-index, mongo-partial, postgres-partial}

\begin{figure}[t!]
    \centering
    \includegraphics[width=0.6\columnwidth]{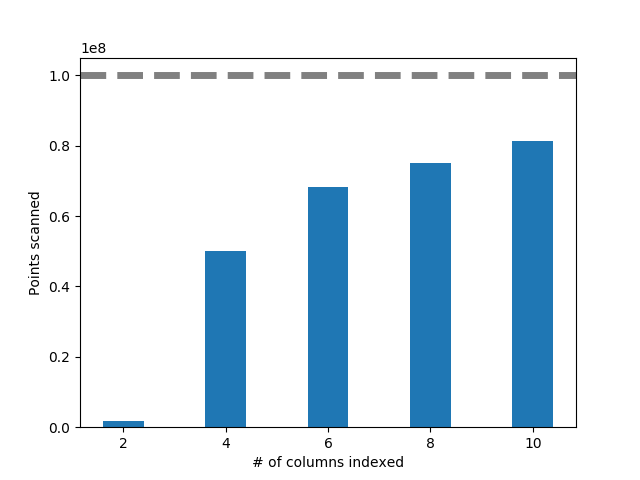}
    \caption{Even on queries with 1\% selectivity, an octree scans almost the entire table (dotted line), a consequence of the curse of dimensionality. The table consists of two independent base columns; every other column is equal to one of the base columns with additive noise drawn from a Pareto distribution with $\alpha = 5$.}
    \label{fig:corr:octree}
\end{figure}

Another alternative is to use a multi-dimensional index to capture both $A$ and $B$. However, \Figure{corr:octree} shows that this approach does not scale past a small number of columns, even if those columns are correlated. This phenomenon is commonly referred to as the \emph{curse of dimensionality} and afflicts multi-dimensional indexes across the board. Moreover, indexing additional columns worsens performance on queries over existing columns in the index, a drawback that secondary indexes do not have.

Both secondary indexes and standard multi-dimensional indexes are unaware of the relationship between $A$ and $B$, and cannot therefore capitalize on it. How might an index take advantage of the correlation to both accelerate query performance and reduce space overhead?
The typical approach to this problem has been to index one of the columns, say $B$, using a separate \emph{host index}, which may or may not be a primary index, and map the values of $A$ onto $B$. At query time, the execution engine consults the mapping to figure out which values of $B$ to scan.

Correlation Maps (CMs)~\cite{cm} do precisely this: for every distinct value (or range of values) in $A$, a CM lists the values (or range of values) of $B$ present in the table. CMs are guaranteed to produce values of $B$ that are a superset of the ranges that need scanning and work well for tight correlations with no outliers. However, they will trigger a scan of an entire primary index page if even a single point lies in it. As a result, their performance degrades substantially in the presence of any outliers~\cite{hermit}.

Since most real datasets can reasonably be expected to have outliers, recent work has developed solutions to handle them without degrading query performance. BHUNT~\cite{bhunt}  finds algebraic constraints, where two columns are related to each other via simple arithmetic operations. However, this class of correlation is extremely restrictive.
Hermit~\cite{hermit} further handles \emph{arbitrary} soft functional dependencies by learning a piecewise linear function that maps $A$ to $B$. In each piecewise component, a manually set error bound around the fitted line determines which records are considered outliers. The outliers are stored in a separate outlier index (a secondary B-Tree), while the inliers are mapped onto ranges of $B$ directly, like a CM, and scanned using the host index.

These prior solutions have two fundamental limitations: they do not generalize to arbitrary types of correlations and are designed for single dimensional host indexes.

 \begin{figure*}
 \centering
    \includegraphics[width=0.24\textwidth]{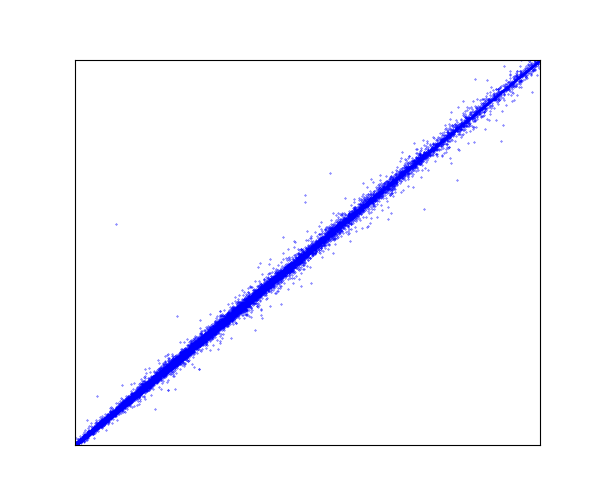}
     \includegraphics[width=0.24\textwidth]{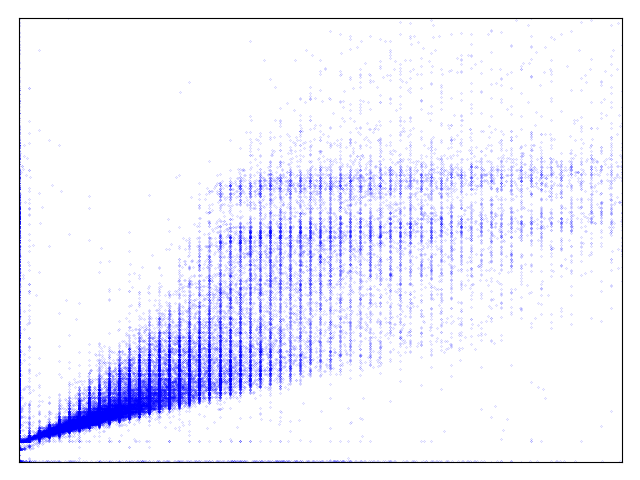}
     \includegraphics[width=0.24\textwidth]{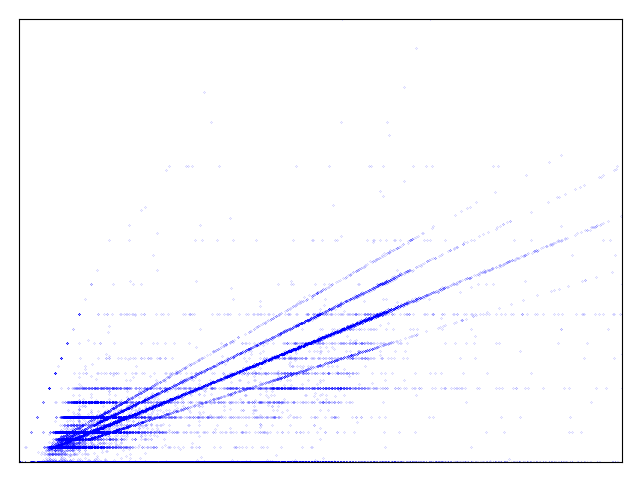}
     \includegraphics[width=0.24\textwidth]{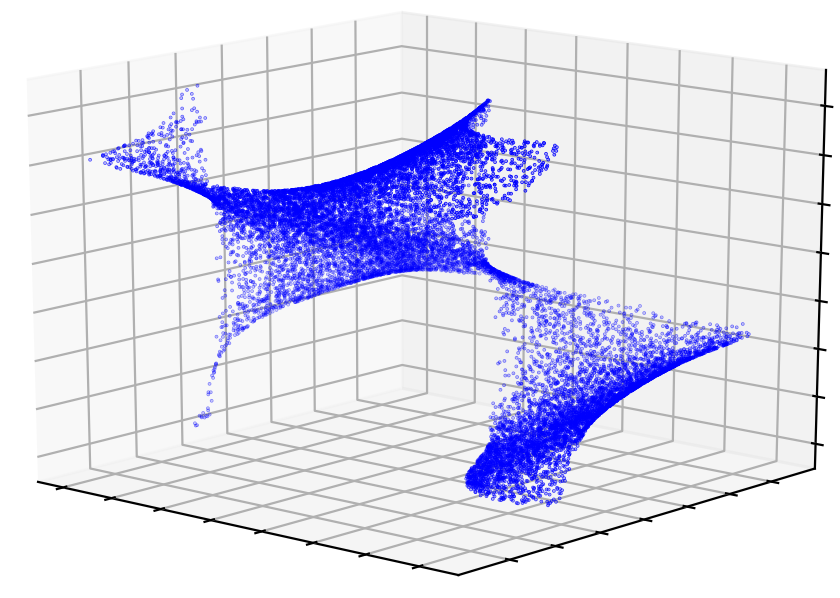}
     \caption{Four types of correlations found in real datasets. Existing solutions target only (a) strong soft-functional dependencies, but not (b) weak correlations, (c) non-functional dependencies, or (d) multi-way correlations.}
     \label{fig:corr:corr_types}
 \end{figure*}

% \jd{Maybe it would be helpful to have a visual depiction of why CM and Hermit do poorly. For example: 6 subfigures arranged in 2 rows and 3 columns. Each subfigure shows the data scanned vs. selected. The columns are CM, Hermit, and Cortex. The first row is a simple correlation where everything performs well. The second row is a non-functional correlation where only cortex does well.}

\NewPara{Correlation Types.} Hermit specifically only caters to \textit{soft functional correlations}, e.g., correlations for which the value of $A$ maps to a small number of values in $B$, plus possible outliers. However, this excludes three important classes of correlations, shown in \Figure{corr:corr_types}:
\begin{CompactEnumerate}
    \item \emph{Weak functional}, in which $a \in A$ may correspond to a large range of $B$ values.
    \item \emph{Non-functional}, in which $a \in A$ may correspond to multiple ranges of $B$ values.
    \item \emph{Multi-way} correlations, in which a column $C$ is correlated to the joint distribution $(A,B)$ but not to either $A$ or $B$ individually. Consider a table with columns on \emph{unit price}, \emph{sale volume}, and \emph{total price}. \emph{Total price} is the product of the other two columns, but is not correlated to either column alone.
    \item Correlations on \emph{categorical} attributes, e.g., ZIP Code and City.
\end{CompactEnumerate}

Prior work is not able to handle this diversity of correlation types: CM can handle both categorical and continuous attributes, but is not equipped to deal with any form of outliers, making it of limited value on real datasets. On the other hand, Hermit handles outliers but is limited in which types of correlations it can index. First, by fitting a functional model to the data, Hermit is not able to capture relationships between categorical attributes. Second, Hermit's piecewise linear model is still a function and cannot fit a non-functional correlation. Third, the assumption of a soft-functional correlation is explicitly hard-coded into its algorithm in the form of a small error tolerance for outliers and a cap on outliers fixed at a small fraction of the table size. These assumptions are usually insufficient at whittling out the right amount of outliers, and tuning the parameters for each correlation is non-trivial.

% \Figure{overview:hermit-fail} shows examples of Hermit's outlier classification on both weak and non-functional correlations. In both cases, Hermit captures some, but not all of the outliers; we will see the effect of this on performance in \Section{eval}. Requiring that Hermit be manually tuned to adapt to each type of configuration is prohibitive; it puts the onus of manual adjustment on the DBA and depends on the complexities of the specific correlation being indexed.

\TheSystem is able to generalize past soft-functional correlations to the three other types listed above. This is made possible through \TheSystem's redefinition of an \textit{outlier}: instead of being defined relative to a fixed model (e.g., linear), outliers are exactly those points that maximize query performance when assigned to the outlier index. This means \TheSystem is \emph{model-free}, since it does not make any assumptions about the relationship between columns; this allows it to handle many more diverse types of correlations.

% \begin{figure}[t!]
% \centering
% \includegraphics[width=0.49\columnwidth]{figures/hermit_outliers_primary_8_5_8.png}
% \includegraphics[width=0.49\columnwidth]{figures/hermit_outliers_primary_8_4_8.png}
% \caption{Outliers (red) determined by Hermit on a non-function and weak linear correlation.}
% \label{fig:overview:hermit-fail}
% \end{figure}

% \ma{Two comments on this section: (1) As Jialin suggested, I think a plot that compares CM, Hermit, and Cortex in terms of what they end up scanning in these complex correlation cases would be more useful. However, I wouldn't both showing the simple correlation case at all. You can use examples of weak and non-functional dependencies or even include a multi-way correlation. (2) Order figures as they are referred to in the text. For example, in Fig. 2: non-functional -> weak -> multi-way.}

\NewPara{Host Index Limitations.}
Hermit assumes the existence of a single-dimensional host index by design: it maps each correlated column to ranges on only one host column. Trying to operate Hermit on a multi-dimensional index yields poor performance, since it does not take into account the index's complex partitioning of attribute values (\Section{eval}). Therefore, Hermit optimizes for only a single dimensional index. However, it may be unlikely that a single-dimensional index will be clustered on a primary key that is well-correlated to other attributes. Hermit thus assumes that the host index is a \emph{secondary} index built on a non-clustered column. This improves Hermit's practicality but at a price: it cannot exploit the locality of records on disk that a primary index offers. As a result, it is fundamentally constrained by the performance of its secondary B-Tree index.

\TheSystem breaks through this limitation by recognizing that multi-dimensional indexes are a solution to the ``uncorrelated primary key'' dilemma. By indexing over many attributes, a multi-dimensional index can provide fast performance on a primary key while also speeding up range queries over other important attributes. Many production systems employ multi-dimensional indexes~\cite{amazon-zorder,oracle-zorder,spark-sql}, and recent work automatically tunes multi-dimensional indexes to retain fast performance on any type of query workload~\cite{flood}. 

Unlike a secondary host index, a (multi-dimensional) primary index induces a sort order on records in physical storage. Pages of the index thus map to contiguous ranges of records in memory, which are much faster to scan than the point lookups incurred by secondary indexes.
Armed with this understanding, \TheSystem targets its design to primary host indexes by mapping a correlated column directly to pages in a primary index. By taking advantage of the primary index, \TheSystem is no longer bound by the performance of the secondary B-Tree.

\section{System Overview}
\label{sec:overview}

\begin{figure*}[th!]
\centering
\includegraphics[width=0.95\linewidth]{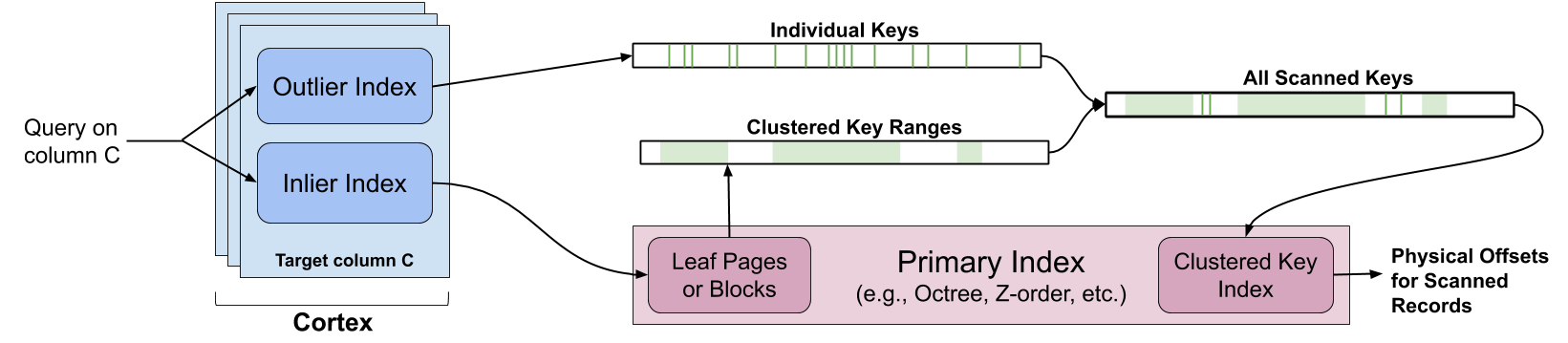}
\caption{A query in Cortex accesses the outlier index for point scans and the inlier index, which maps to pages in the host index, for range scans. After deduplication, keys are translated into physical offsets using the clustered index.}
\label{fig:overview:query_flow}
\end{figure*}

\TheSystem adopts a hybrid design that consists of two key components: a dedicated outlier index using a secondary index structure that supports fast point lookups, and a primary index to perform range scans.
\TheSystem operates in conjunction with the primary index (or \emph{host index}) that already exists on the table, which indexes the \emph{host columns} and dictates how the database is sorted in physical storage.
Each correlation in \TheSystem maps values of the column to be indexed, the \emph{target column}, to a combination of one or more host columns. Note that \TheSystem is capable of supporting secondary host indexes in case the primary index does not contain the desired host columns. 

\subsection{The Host Index}

%todo: cite tsunami
\TheSystem works with any host index that organizes records into pages or blocks. This method of organization is already the natural representation of tree-based or grid-based multi-dimensional indexes, such as octrees, kd-trees, and learned variants~\cite{flood}: the leaves themselves are the pages. Likewise, most single-dimensional indexes partition data into pages of consecutive records. Therefore, \TheSystem is compatible with a wide variety of host index types.

Each record is given a clustered key that is unique and orders records by the host bucket it is in (or their location in the sort order); for example, a clustered key for an octree might consist of the leaf page ID appended with a unique identifier, while for a Z-index, it may simply the Z-score of the record.
A multi-dimensional index then comprises two parts (\Figure{overview:query_flow}): an algorithm, such as an R-tree, that maps a record to a host bucket (which in turn corresponds to a contiguous range of clustered keys), and a clustered index, typically a B-Tree, that maps the clustered keys to physical locations in storage. The use of a clustered key as an intermediate is standard and necessary for efficient inserts and updates~\cite{sql-server-primarykey}.
%sql-server-primarykey: https://docs.microsoft.com/en-us/sql/relational-databases/tables/create-primary-keys

\TheSystem is much more versatile than prior approaches to indexing correlations because it can index any correlation between a target column and \emph{any subset} of the host columns. This lets a host index harness the shared information between correlated columns without having to compromise on which columns to index. For example, in a table with column $A$ correlated to $B$, $D$ correlated to $E$, and $C$ simply a frequently queried column (e.g., timestamp), a host index on $(B, D, C)$ allows \TheSystem to capture both correlations, while still achieving good performance for queries on $C$. This host index would also allow \TheSystem to capture any \textit{multi-way correlations}, e.g., between another column $F$ and $(B,D)$ (or any other subset of host columns).

%For example, consider a table with columns \textit{unit price}, \textit{sale volume}, \textit{total price}, \texit{shipping weight}, and \textit{tax}. \textit{Shipping weight} is correlated with \textit{sale volume}, \textit{tax} is correlated with \textit{total price}, and \emph{total price} is correlated to the combination of \textit{unit price} and \textit{sale volume}, since it is the product of the two.
%Note that \textit{total price} is not individually correlated to either \textit{unit price} or \textit{sale volume}, only to the joint distribution of the two. We can capture all these correlations with a multidimensional host index on \textit{unit price} and \textit{sale volume}, since all other columns are correlated to one or both of these.

\begin{figure}[t!]
    \centering
    \includegraphics[width=0.49\columnwidth]{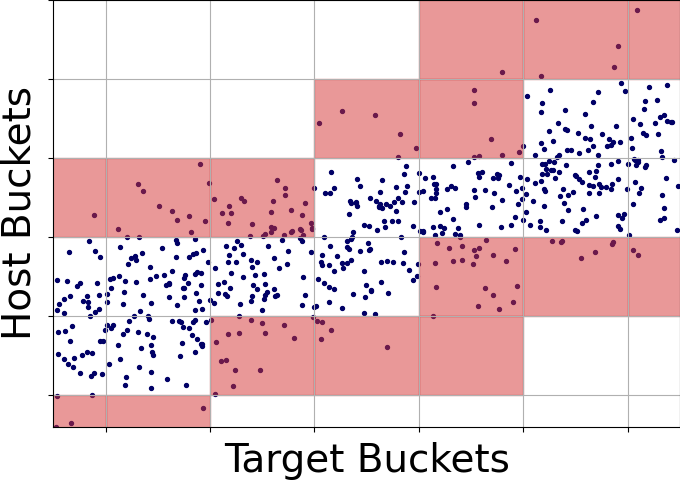}
    \includegraphics[width=0.49\columnwidth]{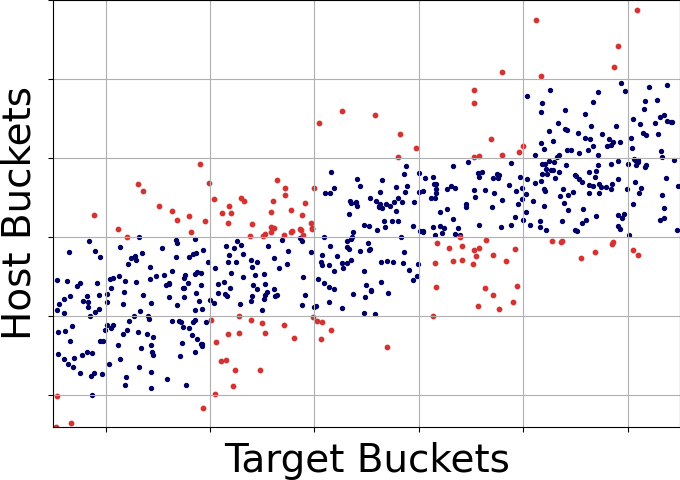}
    \caption{\TheSystem partitions space based on the host and target buckets, and classifies each cell as an inlier or outlier (left). All points inherit the classification of their cell (right).}
    \label{fig:algo:outlier_diagram}
    \vspace{-0.1in}
\end{figure}

\subsection{Indexing}
\label{sec:overview:indexing}
To index a correlation, \TheSystem divides the values of the target column into many \emph{target buckets}, and uses the leaf pages of the host index as a natural partition of the host columns' values into \emph{host buckets} (\Figure{algo:outlier_diagram}). Note that since host buckets cover multiple columns, they may be of different sizes and need not be ordered in any way. 
The target and host buckets partition the table, with each record belonging to a single (target bucket, host bucket) pair, or \emph{cell}.
\TheSystem then runs its custom \emph{stashing algorithm} to determine which cells are \emph{inliers} and \emph{outliers}, detailed in \Section{algo}. 

To use \TheSystem, a database admin (DBA) specifies the correlations they wish to be indexed. For example, they might have a host index on $B, C,$ and $D$ and request that the correlations between $(A,B)$ and $(B, D, E)$ are indexed. \TheSystem consists of three data structures per correlation: an inlier index, outlier index, and correlation tracker. The inlier and outlier indexes participate in query execution, while the correlation tracker maintains the information necessary to handle updates.

\NewPara{Inliers}. The inlier cells are indexed by a structure resembling a Correlation Map~\cite{cm}: for each target bucket $t_i$, the inlier index lists the host buckets $h_j$ for which the cell $(t_i, h_j)$ is an inlier. At query time, if the query has a filter over the target column, \TheSystem scans the matching host buckets, i.e., the union of all host buckets that are inliers for the target buckets that overlap with the filter. Specifically, for a range $R$ on the target column, \TheSystem will scan all points in $\{h\;|\; \exists t \text{ s. t. } (t, h) \text{ is an inlier and } t \cap R \neq \emptyset\}$.
A host bucket $h$ can include points matching multiple target buckets. If at least one of these target buckets intersect the query range $R$, \TheSystem scans all of $h$, potentially scanning unnecessary points.
The inlier index therefore requires a small space overhead to maintain, but it achieves that simplicity at the cost of a potentially higher scan overhead.

\NewPara{Outliers}. For each correlation, each point in an outlier cell is indexed by a separate secondary index structure according to its target column value. Although using a B-Tree as a secondary index is an option, \TheSystem instead stores a list of outliers per target bucket in a sorted list. This approach demands a smaller space overhead and makes deduplication with the inlier index faster at query time (\Section{overview:query_flow}). The outlier index makes the opposite tradeoff from the inlier index: it requires a larger space overhead to maintain but guarantees than the only outliers scanned are those in target buckets touched by the query filter, avoiding superfluous overhead.

\NewPara{Correlation Tracking}. As records are inserted into or deleted from the table, \TheSystem's determination of which cells are inliers and outliers will change. Instead of running the stashing algorithm anew on the entire dataset, it maintains a small data structure to track when individual cells need to be reclassified. The tracker must also handle changes to the host index as a result of insertions and deletions, e.g. if a host bucket is split into multiple children or if many buckets are merged into one. The tracker's operation is described in \Section{algo:online}.

\subsection{Query Flow}
\label{sec:overview:query_flow}

\TheSystem processes a range query over the target column in four steps, outlined in \Figure{overview:query_flow}:
\begin{enumerate}
    \item Query the inlier index to find the host buckets $h_j$ that intersect with the target buckets touched by the range. Each host bucket corresponds to a single range of clustered keys.
    \item Query the outlier index to find the individual clustered keys for points that belong to the same target buckets.
    \item Deduplicate the results from (1) and (2). This is a superset of all points that match the query filter. Deduplication is necessary because inliers and outliers are only guaranteed to be unique \emph{per target bucket}. If a query covers multiple target buckets, the outliers of one bucket may fall within the inlier ranges of another.
    \item Scan the given ranges and individual points, pruning out those that do not match the query filter.
\end{enumerate}

The decision to stash cells in the outlier index versus keeping them in the inlier index affects query performance. Stashed points may be fewer in number than total size of the scanned ranges, but they are random accesses and incur a larger latency depending on the underlying storage structure.
\section{Stashing Algorithm}
\label{sec:algo}

In this section, we describe \TheSystem's algorithm for deciding the \emph{outlier assignment} for a single correlation; that is, which (target bucket, host bucket) cells to classify as outliers. If a cell is an outlier, all the points in that cell are stashed in the outlier index.
We first define the parameters in our formulation~(\Section{algo:parameters}), and then lay out the cost model and \TheSystem's solution to optimize it~(\Section{algo:formulation}),
and then describe how \TheSystem adapts the outlier assignment in the presence of insertions and deletions~(\Section{algo:online}).

\subsection{Parameters}
\label{sec:algo:parameters}

As mentioned in \Section{overview}, the host buckets are simply the pages of the host index, which \TheSystem takes as a given.
\TheSystem's first step is then to divide the values of the target column into target buckets. If the column is a categorical attribute, each bucket is simply a single value. For real-valued attributes or categorical values with too many unique values, the target value range is divided into buckets with equal numbers of points as much as possible (having many points with the same value may make bucket sizes uneven). To determine the number of target buckets $N_t$, we observe that choosing a bucket size that is too small does not generally hurt query speed (but requires more space), while choosing too large a bucket size can significantly degrade performance by incurring a high scan overhead (\Section{eval:target_buckets}).

The combination of host buckets and target buckets partitions the records into disjoint cells. For a clustered host index on a single column, the cells resemble a standard 2D grid. Given a grid, \TheSystem has to determine which grid cells are outliers (\Figure{algo:outlier_diagram}).

With the partition of the target column into target buckets, \TheSystem computes the number of points per cell. We denote the number of points in a cell $(t, h)$ as $|(t, h)|$. Initially, all cells are considered inliers. Classifying a cell $(t, h)$ as an outlier comes with a tradeoff. On one hand, \TheSystem would not have to scan host bucket $h$ if a query only touched $t$; this is especially advantageous when $h$ contains a large number of points outside of $t$. On the other hand, the size of the outlier index increases by $|(t, h)|$, and \TheSystem incurs the performance overhead of doing non-sequential lookups for these points.
Settling on an outlier assignment is therefore a tradeoff between space overhead and query speed. Solving this problem requires some form of storage constraint by the DBA, which \TheSystem allows in one of two forms:
\begin{CompactEnumerate}
\item A hard space limit (equivalently, a maximum number of outliers) that \TheSystem will not exceed when choosing outliers. This is an intuitive constraint but cannot automatically adapt itself to varied correlation types.
\item The relative value of storage and performance, i.e. how much of a performance improvement would you like in exchange for using extra storage?
\end{CompactEnumerate}
\TheSystem supports both formulations, but this section will focus primarily on (2), since we believe that its adaptability to adjust to different strengths of correlations is valuable.

\TheSystem captures the relative value of storage and performance in a parameter $\alpha$, set by the DBA and defined as the percentage improvement in scan time from \TheSystem's initial state (all inliers) that is equal in value to a 1\% increase in storage. This is an intuitive definition: if $\alpha = 10$, \TheSystem would stash 1\% of points as outliers only if doing so improved performance by more than 10\%. A large value of $\alpha$ tends to select very few outliers and thus does not improve performance substantially, while setting $\alpha = 0$ does the opposite, but requires more space for the outlier index.

In addition to $\alpha$, \TheSystem defines a second parameter $\beta$ that encodes the average cost of a non-sequential point lookup, relative to the cost incurred by a point accessed via a range scan. This includes the time required to deduplicate, look up the point's unique clustered key from the secondary index, and fetch the corresponding value from storage using the clustered index.
Unlike $\alpha$, $\beta$ is determined automatically: \TheSystem runs 1000 queries with various selectivities, and fits a linear model of the form:
\[ \text{Query time}=  c_1(\text{\# records in range scans}) + c_2(\text{\# point lookups})+c_3 \]
Then, $\beta \equiv \nicefrac{c_2}{c_1}$. This formulation makes the simplifying assumption that the query time has a linear dependence on the number of scanned points, which we found to be accurate in practice: the correlation coefficient of the resulting fit is $R^2 \gtrsim 0.97$.

Tuning $\beta$ is how \TheSystem adapts itself to varied host indexes and storage formats. For example, a compressed table may hurt point lookup performance more than range scans; this would be reflected in a larger $\beta$. Additionally, if the host index is a secondary index, then $\beta \approx 1$, reflecting that there is no significant advantage to range scans over the host as compared with point looksup in the the outlier stash.

\Table{algo:params} summarizes the parameters described in this section.

\begin{table}[t!]
\centering
\begin{tabular}{|c|c|c|} \hline
\textbf{Parameter} & \textbf{Description} & \textbf{How is it set?} \\ \hline 
$\mathbf{N_t}$ & Number of target buckets & Automatic \\ \hline
$\mathbf{\alpha}$ & Space / speed tradeoff & User-defined \\ \hline
$\mathbf{\beta}$ & Cost of non-sequential lookup & Automatic \\ \hline
\end{tabular}
\caption{The parameters in \TheSystem's optimization problem.}
\label{tab:algo:params}
\vspace{-0.3in}
\end{table}

\subsection{Outlier Assignment}
\label{sec:algo:formulation}

Let $\mathcal{B}$ be the set of all (target bucket, host bucket) cells that contain at least one point. Define an outlier assignment as $\mathcal{A} = \{\mathcal{O}, \mathcal{I}\}$, where $\mathcal{O} \subset \mathcal{B}$ are the outlier cells, and $\mathcal{I} = \mathcal{B} \setminus \mathcal{O}$ are the inliers. We use $\mathcal{I}(t)$ and $\mathcal{I}(h)$ to denote the inlier cells with target bucket $t$ or host bucket $h$, respectively, and likewise for outliers. It will be clear from context which is being used. \TheSystem formulates the cost of an assignment $C(\mathcal{A})$ over a query workload of $N_t$ queries, each of which intersects exactly one target bucket. It then chooses $\mathcal{O}$ to minimize this cost.

$C(\mathcal{A})$ comprises two terms: one for performance overhead and one for space overhead. Space overhead is simply the total number of points in the outlier index:
\[ \text{SO}(\mathcal{A}) = \sum_{(t,h) \in \mathcal{O}} |(t,h)| \]
Performance overhead for a query that intersects target bucket $t$ is the cost of (a) doing a range scan over the inlier points for that query, namely all the points in the host buckets $h$ for which $(t, h)$ is an inlier, and (b) doing point scans on all the outlier points in $t$. In other words:
\[ PO(\mathcal{A}, t) = \sum_{h \in \mathcal{I}(t)} |h| + \beta \sum_{h \in \mathcal{O}(t)} |(t,h)|\]
where $|h|$ refers to the number of points in host bucket $h$.
The total performance overhead is then $PO(\mathcal{A}) = \sum_t PO(\mathcal{A}, t)$.

Before putting everything together, recall that $\alpha$ was defined as the \emph{percentage} decrease in performance overhead that is equivalent in value to a 1\% increase in storage overhead. In other words, $\alpha$ is defined using relative changes in performance and storage overhead, so it needs to be scaled appropriately before assembling the complete cost function. Let $N$ be the number of records in the table, and let $P_0 = PO(\{\emptyset, \mathcal{B}\})$ be the \emph{initial scan overhead}, assuming all cells are inliers. The total cost function is:
\begin{align}
C(\mathcal{A}) &= \sum_t \left[ \sum_{h \in \mathcal{I}(t)} |h| + \beta \sum_{h \in \mathcal{O}(t)} |(t,h)| \right] + \alpha \frac{P_0}{N} \sum_{(t,h) \in \mathcal{O}} |(t,h)| \nonumber \\
&= \sum_t \left[ \sum_{h \in \mathcal{I}(t)} |h| + \left(\beta + \alpha  \frac{P_0}{N}\right) \sum_{h \in \mathcal{O}(t)} |(t,h)| \right]
\label{eq:algo:cost}
\end{align}

Since host buckets do not change as points are stashed, this problem is independent for each target bucket, and can be solved easily: a cell should be an outlier if the first summand in \Equation{algo:cost} is larger than the second, and an inlier otherwise. Specifically, \TheSystem assigns $(t,h)$ to be an outlier if:
\begin{equation}
\left(\beta + \alpha  \frac{P_0}{N}\right) |(t,h)| < |h|
\label{algo:condition}
\end{equation} 
Assigning outliers is therefore linear in the number of cells.

The fact that the solution to this optimization problem is independent for each target bucket is particularly important, since it means that \TheSystem's outlier assignment is correct \emph{regardless of query distribution}. Any query intersecting target bucket $t$ will maximize $C(\mathcal{A})$ using the above assignment, independent of which other buckets it intersects. Note, however, that this is not true if the storage constraint takes the form of a hard cap. In this case, target buckets would need to be weighted by how frequently they are covered, to properly prioritize those that reduce cost the most.

\subsection{Handling Inserts and Deletions}
\label{sec:algo:online}

% Existing solutions DROP a secondary index entirely if the table changes. https://docs.actian.com/actianx/11.1/index.html#page/DatabaseAdmin/Modifying_Secondary_Indexes.htm

\TheSystem supports dynamic updates while maintaining the correctness of its outlier assignment. For brevity, we will focus this section on inserts; deletions and updates operate analogously. 

As records are inserted, \TheSystem's correlation tracker must keep track of when a cell needs to switch from inlier to outlier or vice versa.
In addition, it is possible, even likely, that the host buckets will change: a page may be split into multiple smaller pages if filled past a certain capacity, or merged into another cell if sufficiently empty. \TheSystem needs to handle these changes to the underlying host index as well as changes in cell assignments.
Note that the target buckets remain largely unchanged, except for when points appear outside the existing range of the target column. If a target bucket becomes disproportionally crowded due to skewed inserts, it may be split into two. However, this isn't required for correctness of the outlier assignment. 

\begin{figure}[t!]
\includegraphics[width=0.9\columnwidth]{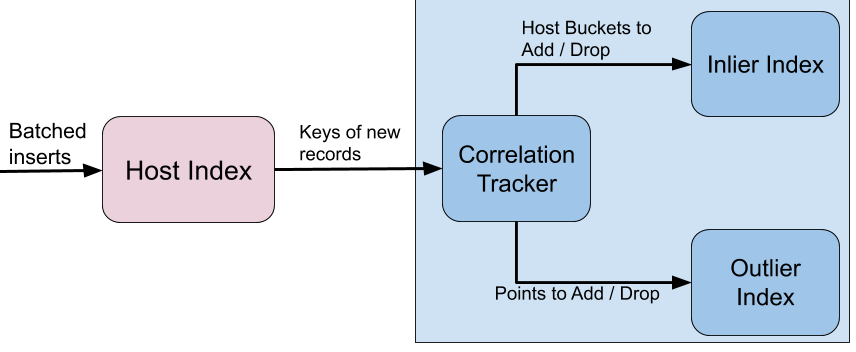}
\caption{\TheSystem's insertion flow.}
\label{fig:algo:insertions}
\vspace{-0.2in}
\end{figure}

\Figure{algo:insertions} illustrates the workflow for insertions. \TheSystem passes all insertions through the host index, and requires that the host index return the clustered keys of the new points, as well as a pointer to (or ID of) the host bucket it was added to.
With this information, \TheSystem updates its internal count of points in each host bucket and determines whether any cells in that host bucket have a new assignment. Any of the following may happen:

\NewPara{The outlier assignment is unchanged.}
Some inserted points belong to inlier cells, and no action needs to be taken for them. The inlier index will visit them when it scans the corresponding host bucket. Points that fall into outlier buckets must be inserted into the outlier index. Unlike a traditional secondary index, only a fraction of points typically need to be inserted.

\NewPara{ A cell $(t,h)$ changes from outlier to inlier.} \TheSystem adds $h$ to the entry for $t$ in the inlier index. In addition, the tracker signals to the outlier index that the points lying within $t$'s value range and $h$'s clustered key range must be removed. This operation scales as the number of points in target bucket $t$ that are outliers.

\NewPara{A cell $(t,h)$ changes from inlier to outlier.} This is the most costly operation. Since \TheSystem does not individually keep track of which records are outliers, it must scan $h$ in the underlying table to find the points that are also in $t$. Since the table is clustered by $h$, this scan can access a host bucket exactly without false positives.

\NewPara{A target bucket is split or merged.} This case occurs least frequently of the four. \TheSystem must recompute the cells for all points in the original target bucket(s) and reassign them all. Since there are often a large number of target buckets, this operation typically touches only a small fraction of the data.

If the inserts have the same distribution as existing records, cells switching between inlier and outlier will be rare. Inserts in order of clustered key are also fast since they will update fewer host buckets.
We evaluate \TheSystem's insert performance in \Section{eval:inserts}.

\section{Evaluation}
\label{sec:eval}

\begin{table*}[t!]
\centering
\small
\begin{tabular}{llllllp{5cm}}
\toprule
 & \textbf{Records} & \textbf{Columns} & \textbf{Correlations} & \textbf{Host Index} & \textbf{Host Columns} & \textbf{Target Columns}
 \\ \midrule
\multirow{2}{*}{\textbf{Stocks}} &
\multirow{2}{*}{165M} &
\multirow{2}{*}{7} &
\multirow{2}{*}{Linear functional} &
1-D Clustered & Daily Open & Daily High, Daily Low, Daily Close \\ \cmidrule{5-7}
\multirow{2}{*}{} & \multirow{2}{*}{} &
\multirow{2}{*}{} & \multirow{2}{*}{} &
Octree & Date, Daily Open &
Daily High, Daily Low, Daily Close \\ \midrule
\multirow{2}{*}{\textbf{Chicago Taxi}} &
\multirow{2}{*}{194M} &
\multirow{2}{*}{9} &
\multirow{2}{*}{\begin{tabular}[t]{@{}l@{}} Weak functional \\ Non-functional \\ Multi-way \end{tabular}} &
 1-D Clustered & Total Fare &
Distance, Metered Fare, Tips  \\ \cmidrule{5-7}
\multirow{2}{*}{} &
\multirow{2}{*}{} &
\multirow{2}{*}{} &
\multirow{2}{*}{} &
Octree &
Start time, Duration, Total Fare &
End Time, Distance, Metered Fare, Tips \\ [-0.5em] \\
\midrule
\textbf{WISE} & 198M & 15 & \multirow{2}{*}{\begin{tabular}{@{}l}
Weak functional \\ Non-functional \\ Multi-way \end{tabular}} &
1-D Clustered & W1 Magnitude & W1 $\sigma$, W1 SNR \\ \cmidrule{5-7}
\multirow{2}{*}{} &
\multirow{2}{*}{} &
\multirow{2}{*}{} &
\multirow{2}{*}{} &
Octree &
RA, Declension, W1 Magnitude &
Galactic Lon/Lat, Ecliptic Lon/Lat, W1 Magnitude, W1 $\sigma$, W1 SNR
\\ \bottomrule
\end{tabular}
\caption{Three real datasets used in our evaluation. We evaluate each dataset on both a single dimensional clustered index and host index; each setting uses a different set of host and target columns.}
\label{tab:eval:datasets}
\vspace{-0.1in}
\end{table*}

This section benchmarks \TheSystem against various other correlation index baselines, as well as a secondary B-Tree index. Our results demonstrate that across our real datasets, and across both single and multi-dimensional host indexes, \TheSystem can outperform the B-Tree while occupying $5-70\times$ less space. At the same time, \TheSystem outperforms prior approaches by $2-8\times$ on a variety of datasets and workloads.

\subsection{Implementation}

%sql-server-impl: https://dl.acm.org/doi/abs/10.1145/1989323.1989448
\TheSystem is implemented in C++. We run all experiments on an AWS r5a.4xlarge instance, with 16 2.2GHz AMD EPYC 7571 processors and 64GB DRAM. 
All experiments use 64-bit integer-valued attributes. Floating point values are multiplied by the smallest power of 10 that makes them integers, and categorical variables are assigned integer values using a dictionary encoding~\cite{sql-server-impl}. Unless indicated otherwise, each query in our query workloads has a range filter over a single target column, and returns the IDs of the records that match the filter\footnote{We find that \TheSystem maintains similar relative performance to the baselines regardless of which aggregation is used.}. Each query range is sampled uniformly at random from the value range of the target column. Each query workload has a target selectivity $s = \frac{\text{points in result}}{\text{table size}}$. Target selectivities are approximate, and the actual selectivity of queries in the workload lies within $[\frac{s}{2}, 2s]$. To clear ambiguity, we say that workloads with larger $s$ have \emph{higher selectivity} than those with smaller values of $s$.

%\ma{isn't larger $s$ the opposite of what it means to be more selective? anyway you're defining it but use whatever is standard terminology in the DB community}. \vikram{I searched some other references and found higher s => higher selectivity to be standard.}

\TheSystem runs on a custom in-memory column store with a bit-packing compression scheme, in which the value in each compression block is encoded as a low bit-width offset from the block's base value. With this setup, we measure $\beta = 17.88$ (the relative cost of a point lookup vs. a range scan; see \S\ref{sec:algo:parameters}) with correlation coefficient of $R^2 = 0.97$.
We use this column store for all baselines:
\begin{enumerate}
\item \emph{Full Scan} scans all records in the table.
\item \emph{Secondary} builds a B-Tree secondary index for every column in the query workload. The B-Tree returns physical pointers to all matching records. The B-Tree we use is cache-optimized with 256-byte nodes~\cite{cpp-btree}.
\item \emph{CM} builds a Correlation Map for every column in the query workload, adapting it to a multi-dimensional setting by mapping buckets along the target dimension to the minimal set of host buckets that are guaranteed to contain all matching points. The CM blocks along the target dimension are chosen identically to \TheSystem's target buckets.
\item \emph{Hermit} implements the TRS-Tree from Hermit~\cite{hermit}, which consists of four manually set parameters: the error bound, which corresponds to an inlier confidence interval for each piecewise linear segment, a maximum outlier fraction, and the maximum depth and fanout of the TRS Tree. We use the recommended parameters from~\cite{hermit}: 2, 0.1, 10, and 8, respectively. Since Hermit can only map the target column to a single host column, in situations with a multi-dimensional host index, we choose the host column with the highest correlation to the target column being indexed.
\item \emph{\TheSystem} is our solution for $\alpha = 0.2, 0.5, 1, 2, 5$. These values of $\alpha$ cover the spectrum from minimizing storage to maximizing performance. Since we evaluate workloads down to 0.01\% selectivity, we choose 20000 target buckets for each correlation. \Section{eval:target_buckets} examines the impact of bucketing on performance.
\end{enumerate}

\begin{figure}[t!]
\centering
\includegraphics[width=\columnwidth]{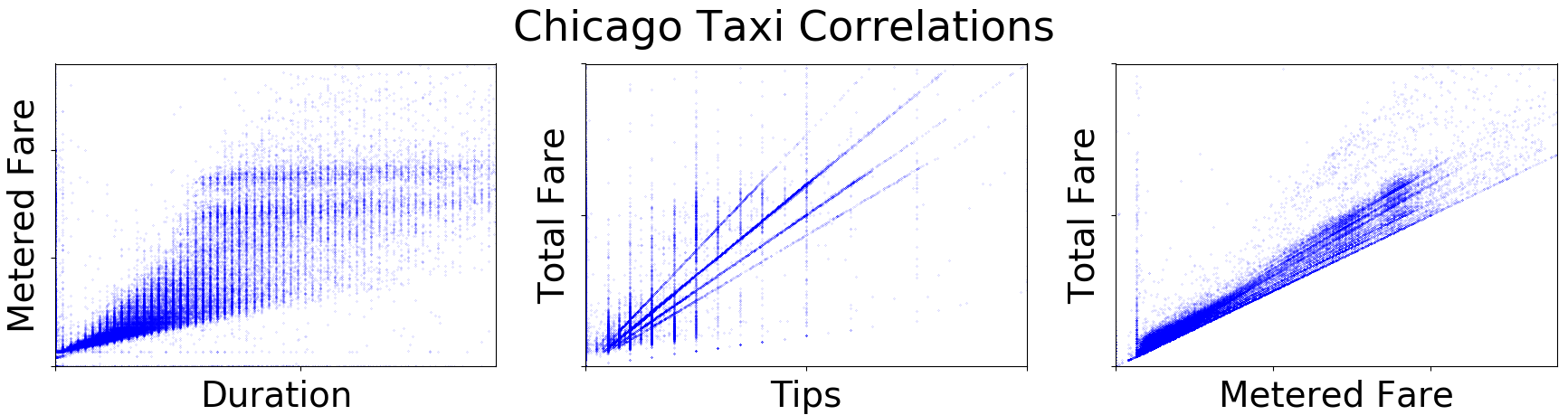}
\caption{Examples of correlations in the Chicago Taxi dataset. Some outliers may not be shown.}
\label{fig:eval:taxi:corr}
\vspace{-0.1in}
\end{figure}

\begin{figure}[t!]
\centering
\includegraphics[width=\columnwidth]{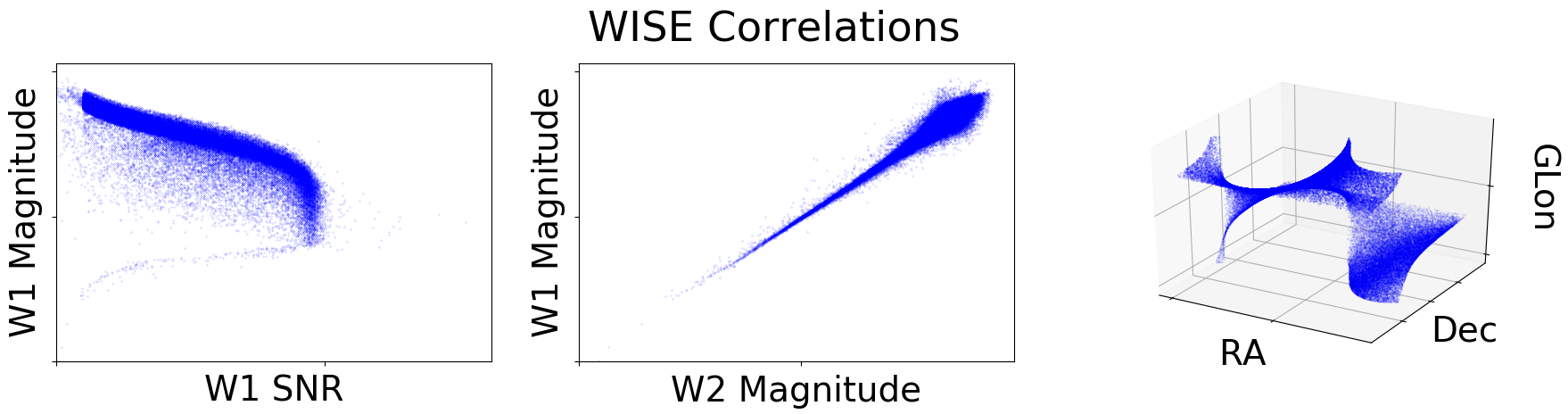}
\caption{Examples of correlations in the WISE dataset. Some outliers may not be shown.}
\label{fig:eval:wise:corr}
\vspace{-0.1in}
\end{figure}

\begin{figure*}[t!]
    \centering
    \includegraphics[width=0.5\textwidth]{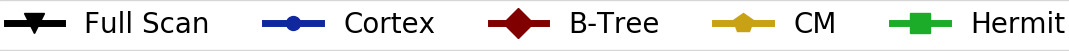} \\
    \subfloat{
        \includegraphics[width=0.25\linewidth]{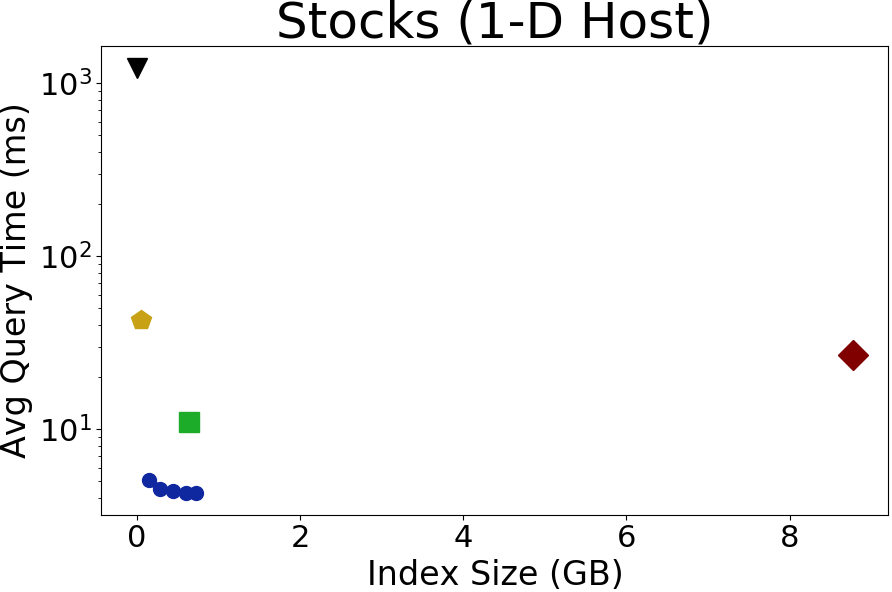}
    }\hspace{7mm}
    \subfloat{
        \includegraphics[width=0.25\linewidth]{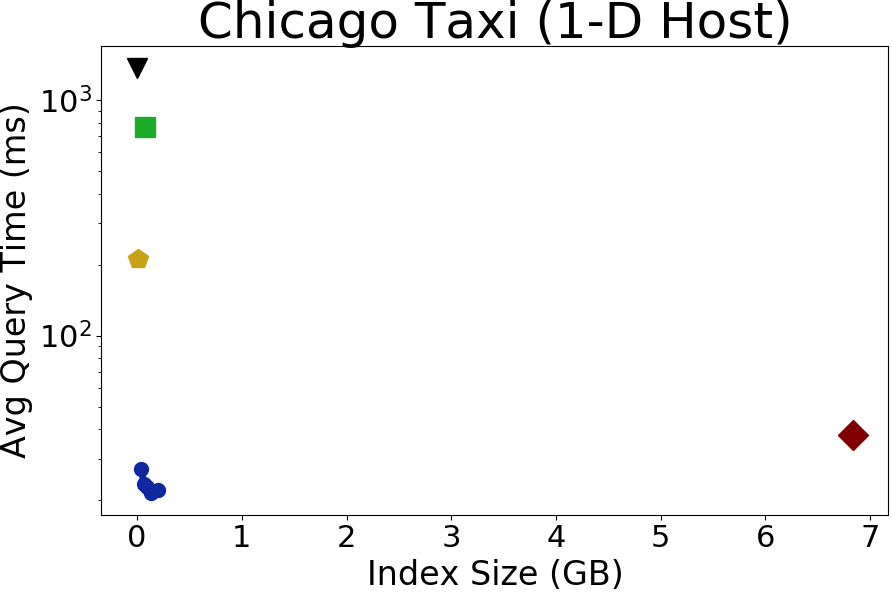}
    }\hspace{7mm}
    \subfloat{
        \includegraphics[width=0.25\linewidth]{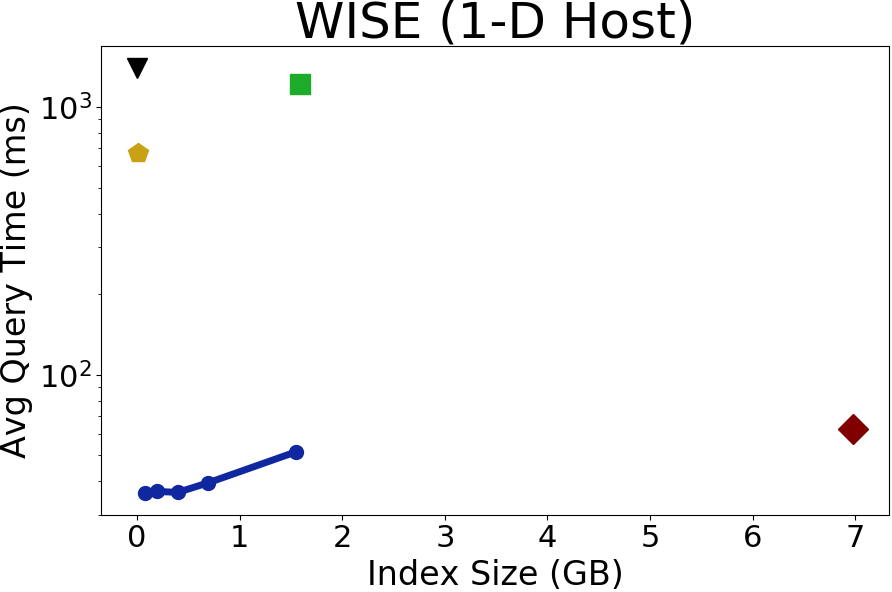}
    }

    \caption{Performance of \TheSystem and baselines on three datasets using a clustered 1-D host index (0.1\% selectivity). Note the log scale.}
    \label{fig:eval:punchline:btree}
\end{figure*}

\begin{figure*}[t!]
    \centering
    \includegraphics[width=0.5\textwidth]{figures/pareto_legend.png} \\
    \subfloat{
        \includegraphics[width=0.25\linewidth]{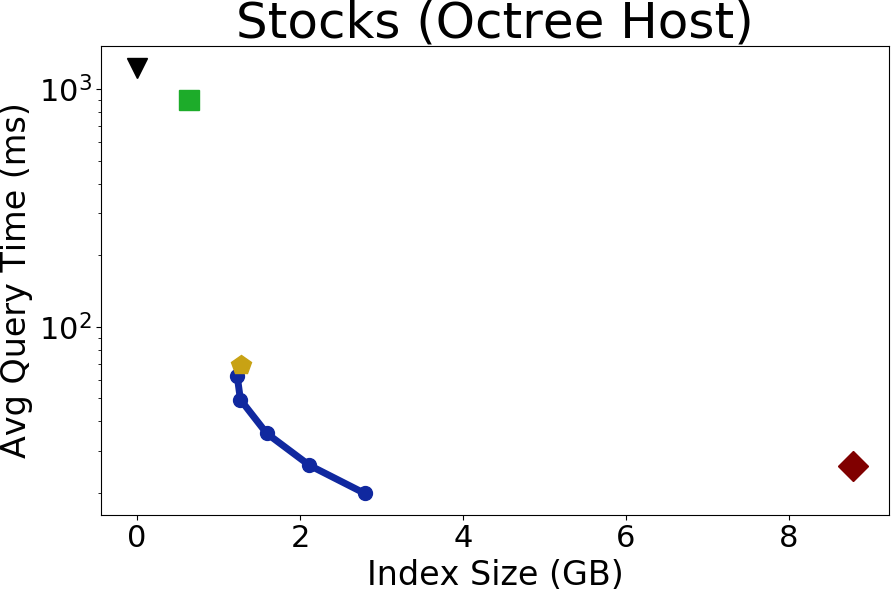}
    }\hspace{7mm}
    \subfloat{
        \includegraphics[width=0.25\linewidth]{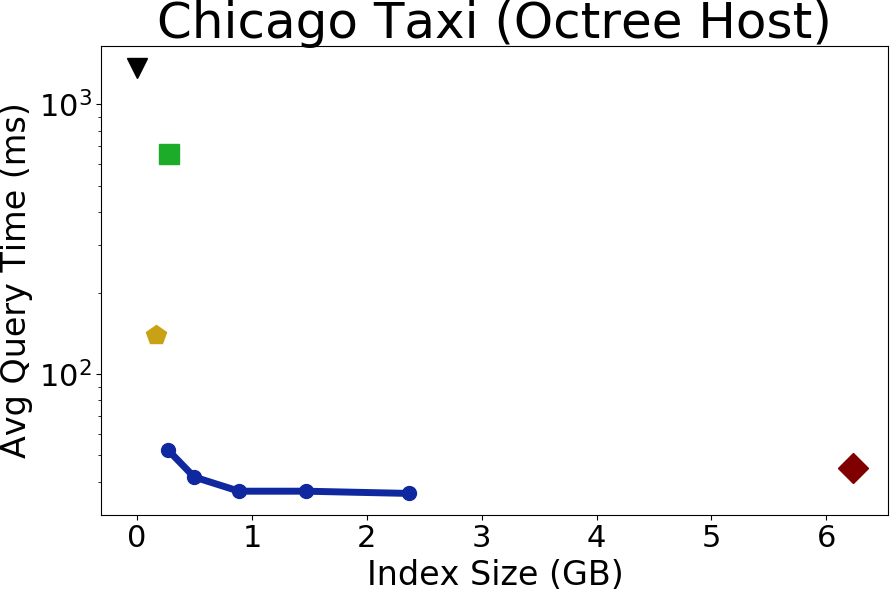}
    }\hspace{7mm}
    \subfloat{
        \includegraphics[width=0.25\linewidth]{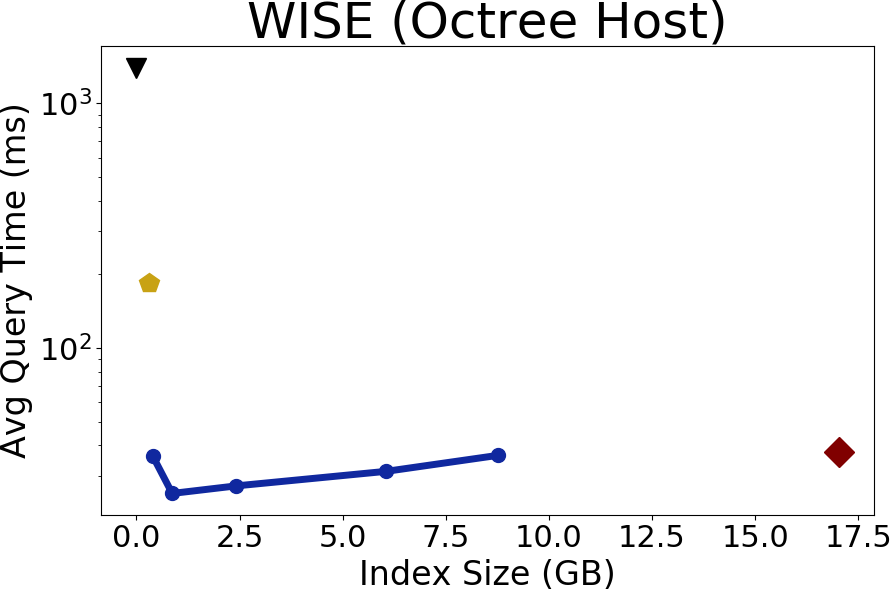}
    }
    \caption{Performance of \TheSystem and baselines on three datasets, using an Octree host index (0.1\% selectivity). Note the log scale.}
    \label{fig:eval:punchline:octree}
\end{figure*}

We evaluate \TheSystem and all baselines on two host indexes:
\begin{enumerate}
\item \emph{Clustered 1-D} is a one-dimensional primary index sorted by a single column that performs a binary search to look up a range. \TheSystem splits the host column into up to 100,000 host buckets. The clustered key is the column value, plus a unique ID in case of duplicates.
\item \emph{Octree} is a hyperoctree in two or more dimensions, with a maximum leaf page size of 10,000. The clustered key for a record is its page ID with a counter to uniquely identify it within the page. We choose an Octree, since results from~\cite{flood} suggest it is the most performant of traditional multi-dimensional index structures.
\end{enumerate}
In both cases, the clustered key index is a cache-optimized B-Tree, similar to the \emph{Secondary} baseline.
 Note that each experiment includes randomly shuffled queries over only the \emph{correlated} (i.e. target) dimensions, but \emph{excludes} queries over the host columns. Therefore, all queries utilize the correlation index, which focuses the evaluation on the performance of the baselines alone.

\subsection{Datasets}
\label{sec:eval:datasets}

We evaluate \TheSystem on several real large-scale datasets with a variety of correlation types, summarized in \Table{eval:datasets}. Note that host columns vary based on host index and dataset.

\NewPara{Stocks} is a dataset with 165M rows comprising daily statistics for over 70k equities listed on various international exchanges, the earliest of which has data from 1962~\cite{tiingo}. The columns are date, stock ticker, daily high, daily low, open, close, and volume. The high, low, open, and close attributes exhibit a tight soft-functional linear correlation, since the price does not usually fluctuate substantially (more than a few percentage points) over a single day. \Figure{corr:corr_types} showed a soft-functional correlation from this dataset.

\NewPara{Chicago Taxi} is a dataset with data on 194M taxi trips in Chicago~\cite{chicago-taxi}. Columns include start time, end time, duration, distance, metered fare, tips, tolls, and total fare. This dataset covers a wide range of correlations (\Figure{eval:taxi:corr}): the metered and total fares are well correlated with the combination of (duration, distance), but have a very weak relationship with each individually. This relationship is not simply linear: for example, longer trips tend to use a flat fare. Tips have a non-functional correlation with total fare (since passengers tend to choose one of several fixed percentages), while total fare and metered fare are moderately correlated.

\NewPara{WISE} samples 198M astronomical objects from the NASA Wide-Field Infrared Survey Explorer~\cite{wise-nasa}. Among its 15 attributes are coordinates for each object in 3 different coordinate systems: Right Ascension (RA) and Declension (Dec), galactic longtitude and latitude (GLon/GLat), and ecliptic longtitude and latitude (ELon/ELat). The two coordinates in any one coordinate system together determine each coordinate in every other system, making this a strong multi-way correlation. In addition to coordinates, the dataset contains two photometric measurements, $W1$ and $W2$, along with 2 different error measurements for each. \Figure{eval:wise:corr} shows examples of the non-functional, weak functional, and multi-way correlations in the WISE dataset. Unlike the Chicago Taxi dataset, where many complex correlations still had strong linear components, the correlations in WISE are more diffuse. NULL values are present and are assigned a large negative value. These skew Hermit's linear model by producing a large outlier tolerance, so we automatically assign them as outliers in Hermit, and disregard them when building the piecewise linear model.

\subsection{Performance Comparison}
\label{sec:eval:punchline}

\Figure{eval:punchline:btree} and \Figure{eval:punchline:octree} show the tradeoff between memory footprint and query performance for \TheSystem and the other baselines, on a query workload with 0.1\% selectivity. Note that the correlated target columns being indexed (and therefore the query workload) are different for single and multi-dimensional host indexes (see \Table{eval:datasets}); as a result, the size of an index will differ between the two settings. \TheSystem's memory footprint decreases as $\alpha$ increases, so in each graph, $\alpha = 0.2$ corresponds to the rightmost point, while $\alpha=5$ is the leftmost. All datasets demonstrate that, on queries with 0.1\% selectivity, \TheSystem establishes the Pareto frontier of the size-speed tradeoff; in other words, any other approach operating at the same performance as \TheSystem requires more space. In particular, across the three datasets, \TheSystem can \emph{match} or \emph{outperform} a B-Tree's performance while using $17.5-71\times$ and $5.5-7\times$ less space on single and multi-dimensional host indexes, respectively. At the same time, \TheSystem outperforms the next best correlation index by between $2-18\times$, with a wider margin ($6.4-18\times$) on the weaker correlations in the Chicago Taxi and WISE datasets. Note that we don't include Hermit in the evaluation of the WISE dataset with an Octree host, since it cannot fit the multi-way correlations. For example, GLon is nearly independent to any other single attribute. When Hermit tries to fit this correlation, it hits its max depth, occupies more memory than the secondary index, and takes an inordinate amount of time.

% \ma{Is there actually any advantage to the multi-dimensional host index over the 1-D index? Why does the size of the secondary index change in Fig. 9 and 10?}

% \ma{I think you should use different texture for range and point in Figures 11 and 12. For example, in Figure 11, I can't tell what is range or point. (it can be figured out but make it easy for the reader)}

 \begin{figure}[t!]
     \centering
     \includegraphics[width=\columnwidth]{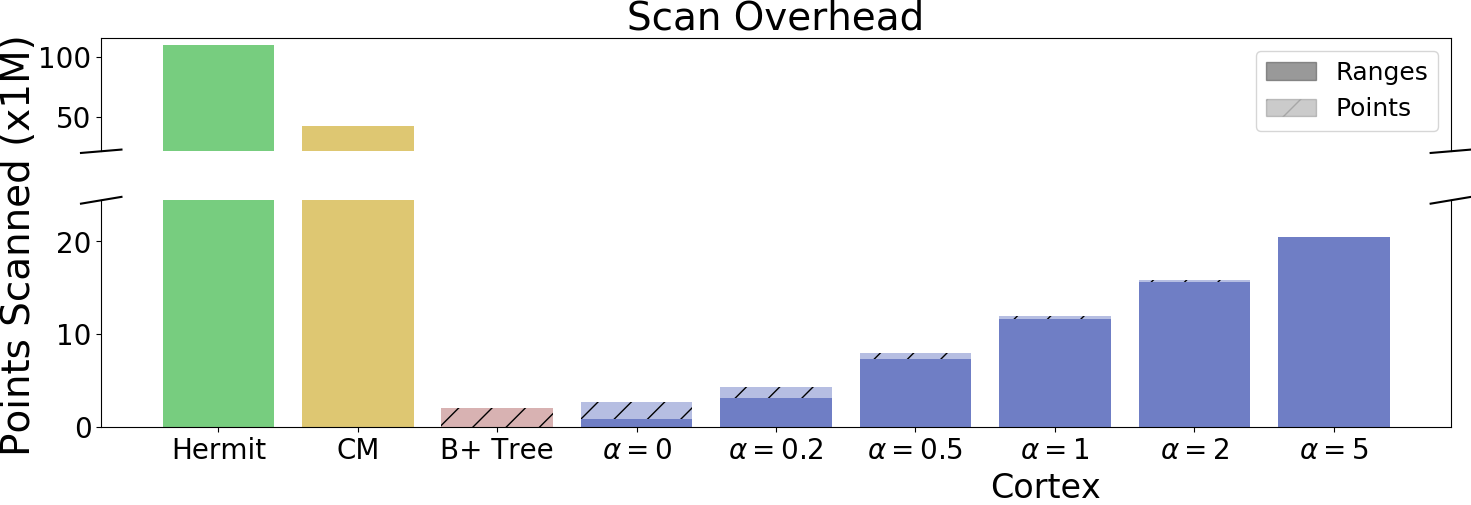}
     \caption{\TheSystem scans a smaller range than CM or Hermit, with an octree host index on Chicago Taxi (1\% selectivity).}
     \label{fig:eval:alpha_points}
 \end{figure}

On the Stocks dataset, \TheSystem can outperform Hermit and CM on a single-dimensional host index, despite their specializing in these types of soft-functional correlations. It's performance improvement is less substantial on the Octree host, which we investigate further below.
The gap between \TheSystem and these previous solutions grows wider on the Chicago Taxi and WISE datasets, which have weaker correlations.
CM suffers by not being able to account for the outliers on the fringes of these weaker correlations, which results in its scanning a large range (\Figure{eval:alpha_points}). On the other hand, Hermit's outlier threshold filters out almost the entire dataset, which we investigate further below. A tight bound on outliers works well on strong correlations but quickly degrades in the presence of weaker ones.

We now take a deeper look at how \TheSystem's performance varies based on multiple parameters: query selectivity, the parameter $\alpha$, the choice of host index, and the strength of the correlation.

\begin{table*}[t!]
\setlength{\aboverulesep}{0pt}
\setlength{\belowrulesep}{0pt}
\centering
\begin{tabular}{|c||cccc||cccc||cccc|} \toprule
\multirow{2}{*}{\textbf{Baseline}} & \multicolumn{4}{c||}{\textbf{Stocks}} & \multicolumn{4}{c||}{\textbf{Chicago Taxi}} & \multicolumn{4}{c|}{\textbf{WISE}} \\ 
\multirow{3}{*}{} & \textbf{0.01\%} & \textbf{0.1\%} & \textbf{1\%} & \textbf{5\%} & \textbf{0.01\%} & \textbf{0.1\%} & \textbf{1\%} & \textbf{5\%} & \textbf{0.01\%} & \textbf{0.1\%} & \textbf{1\%} & \textbf{5\%} \\ \hline
Full Scan& 0.002 \cellcolor[rgb]{0.701961,0.152941,0.062745} & 0.022 \cellcolor[rgb]{0.701961,0.152941,0.062745} & 0.173 \cellcolor[rgb]{0.772975,0.354771,0.286066} & 0.878 \cellcolor[rgb]{0.983189,0.952220,0.947133} & 0.004 \cellcolor[rgb]{0.701961,0.152941,0.062745} & 0.028 \cellcolor[rgb]{0.701961,0.152941,0.062745} & 0.220 \cellcolor[rgb]{0.803952,0.442810,0.383480} & 0.974 \cellcolor[rgb]{0.996593,0.990317,0.989286} & 0.005 \cellcolor[rgb]{0.701961,0.152941,0.062745} & 0.045 \cellcolor[rgb]{0.701961,0.152941,0.062745} & 0.193 \cellcolor[rgb]{0.787382,0.395716,0.331371} & 0.683 \cellcolor[rgb]{0.950705,0.859899,0.844981} \\
Correlation Map& 0.482 \cellcolor[rgb]{0.905657,0.731868,0.703317} & 0.635 \cellcolor[rgb]{0.941231,0.832972,0.815187} & 1.843 \cellcolor[rgb]{0.826064,0.925010,0.815649} & 3.495 \cellcolor[rgb]{0.644093,0.846555,0.622782} & 0.098 \cellcolor[rgb]{0.701961,0.152941,0.062745} & 0.179 \cellcolor[rgb]{0.777063,0.366389,0.298921} & 0.495 \cellcolor[rgb]{0.909007,0.741389,0.713851} & 1.427 \cellcolor[rgb]{0.898906,0.956414,0.892852} & 0.015 \cellcolor[rgb]{0.701961,0.152941,0.062745} & 0.093 \cellcolor[rgb]{0.701961,0.152941,0.062745} & 0.348 \cellcolor[rgb]{0.863420,0.611826,0.570492} & 1.229 \cellcolor[rgb]{0.941350,0.974714,0.937838} \\
Hermit& 1.842 \cellcolor[rgb]{0.826274,0.925100,0.815871} & 2.463 \cellcolor[rgb]{0.743653,0.889479,0.728303} & 2.377 \cellcolor[rgb]{0.753757,0.893835,0.739012} & 3.402 \cellcolor[rgb]{0.651753,0.849858,0.630900} & 0.005 \cellcolor[rgb]{0.701961,0.152941,0.062745} & 0.049 \cellcolor[rgb]{0.701961,0.152941,0.062745} & 0.350 \cellcolor[rgb]{0.864210,0.614070,0.572976} & 1.069 \cellcolor[rgb]{0.980944,0.991784,0.979803} & 0.006 \cellcolor[rgb]{0.701961,0.152941,0.062745} & 0.051 \cellcolor[rgb]{0.701961,0.152941,0.062745} & 0.250 \cellcolor[rgb]{0.820699,0.490407,0.436145} & 0.648 \cellcolor[rgb]{0.943917,0.840607,0.823634} \\
\textbf{\TheSystem} & 2.934 \cellcolor[rgb]{0.693818,0.867994,0.675484} & 6.149 \cellcolor[rgb]{0.483392,0.777271,0.452457} & 5.351 \cellcolor[rgb]{0.522948,0.794325,0.494382} & 6.347 \cellcolor[rgb]{0.474397,0.773393,0.442924} & 1.336 \cellcolor[rgb]{0.917672,0.964505,0.912742} & 1.654 \cellcolor[rgb]{0.856897,0.938303,0.848328} & 2.449 \cellcolor[rgb]{0.745209,0.890150,0.729952} & 3.012 \cellcolor[rgb]{0.686371,0.864783,0.667591} & 0.956 \cellcolor[rgb]{0.994228,0.983597,0.981850} & 1.722 \cellcolor[rgb]{0.845465,0.933374,0.836211} & 1.221 \cellcolor[rgb]{0.943235,0.975526,0.939836} & 2.484 \cellcolor[rgb]{0.741191,0.888418,0.725694} \\
\bottomrule
\end{tabular}
\caption{Multiplicative speedup of \TheSystem and other baselines compared to an optimized secondary B-Tree index, across a range of query selectivities. All experiments use a single-dimensional clustered host index.}
\label{tab:eval:punchline:selectivity}
\vspace{-0.1in}
\end{table*}

\begin{table*}[t!]
\setlength{\aboverulesep}{0pt}
\setlength{\belowrulesep}{0pt}
\centering
\begin{tabular}{|c||cccc||cccc||cccc|} \toprule
\multirow{2}{*}{\textbf{Baseline}} & \multicolumn{4}{c||}{\textbf{Stocks}} & \multicolumn{4}{c||}{\textbf{Chicago Taxi}} & \multicolumn{4}{c|}{\textbf{WISE}} \\ 
\multirow{3}{*}{} & \textbf{0.01\%} & \textbf{0.1\%} & \textbf{1\%} & \textbf{5\%} & \textbf{0.01\%} & \textbf{0.1\%} & \textbf{1\%} & \textbf{5\%} & \textbf{0.01\%} & \textbf{0.1\%} & \textbf{1\%} & \textbf{5\%} \\ \hline
Full Scan& 0.003 \cellcolor[rgb]{0.701961,0.152941,0.062745} & 0.021 \cellcolor[rgb]{0.701961,0.152941,0.062745} & 0.171 \cellcolor[rgb]{0.771260,0.349896,0.280672} & 0.903 \cellcolor[rgb]{0.986795,0.962471,0.958475} & 0.006 \cellcolor[rgb]{0.701961,0.152941,0.062745} & 0.033 \cellcolor[rgb]{0.701961,0.152941,0.062745} & 0.214 \cellcolor[rgb]{0.800608,0.433306,0.372964} & 1.004 \cellcolor[rgb]{0.998889,0.999521,0.998823} & 0.006 \cellcolor[rgb]{0.701961,0.152941,0.062745} & 0.027 \cellcolor[rgb]{0.701961,0.152941,0.062745} & 0.176 \cellcolor[rgb]{0.775456,0.361822,0.293868} & 0.823 \cellcolor[rgb]{0.974811,0.928409,0.920786} \\
Correlation Map& 0.112 \cellcolor[rgb]{0.717020,0.195741,0.110102} & 0.376 \cellcolor[rgb]{0.873387,0.640154,0.601837} & 0.525 \cellcolor[rgb]{0.916542,0.762804,0.737547} & 1.500 \cellcolor[rgb]{0.884598,0.950246,0.877687} & 0.148 \cellcolor[rgb]{0.752660,0.297034,0.222181} & 0.319 \cellcolor[rgb]{0.852114,0.579692,0.534937} & 0.911 \cellcolor[rgb]{0.987972,0.965816,0.962176} & 2.476 \cellcolor[rgb]{0.742084,0.888803,0.726640} & 0.089 \cellcolor[rgb]{0.701961,0.152941,0.062745} & 0.202 \cellcolor[rgb]{0.793164,0.412149,0.349554} & 0.566 \cellcolor[rgb]{0.926415,0.790864,0.768594} & 2.140 \cellcolor[rgb]{0.783596,0.906700,0.770637} \\
Hermit& 0.005 \cellcolor[rgb]{0.701961,0.152941,0.062745} & 0.029 \cellcolor[rgb]{0.701961,0.152941,0.062745} & 0.168 \cellcolor[rgb]{0.769311,0.344358,0.274545} & 0.820 \cellcolor[rgb]{0.974242,0.926792,0.918996} & 0.010 \cellcolor[rgb]{0.701961,0.152941,0.062745} & 0.068 \cellcolor[rgb]{0.701961,0.152941,0.062745} & 0.356 \cellcolor[rgb]{0.866218,0.619778,0.579291} & 1.500 \cellcolor[rgb]{0.884699,0.950289,0.877795}  &  &  &  & \\
\textbf{\TheSystem} & 0.367 \cellcolor[rgb]{0.870138,0.630918,0.591617} & 0.726 \cellcolor[rgb]{0.958478,0.881990,0.869424} & 0.778 \cellcolor[rgb]{0.967485,0.907589,0.897749} & 1.661 \cellcolor[rgb]{0.855738,0.937803,0.847099} & 1.919 \cellcolor[rgb]{0.814660,0.920093,0.803562} & 2.043 \cellcolor[rgb]{0.796809,0.912397,0.784642} & 2.962 \cellcolor[rgb]{0.691178,0.866855,0.672686} & 5.232 \cellcolor[rgb]{0.529319,0.797072,0.501135} & 0.718 \cellcolor[rgb]{0.957037,0.877895,0.864893} & 1.372 \cellcolor[rgb]{0.909954,0.961178,0.904563} & 1.172 \cellcolor[rgb]{0.954773,0.980501,0.952065} & 2.280 \cellcolor[rgb]{0.765630,0.898954,0.751596} \\
\bottomrule
\end{tabular}
\caption{Multiplicative speedup of \TheSystem and other baselines compared to an optimized secondary B-Tree index, across a range of query selectivities. All experiments use an octree host index.}
\label{tab:eval:punchline:selectivity_octree}
\vspace{-0.1in}
\end{table*}

\begin{figure}[t!]
\includegraphics[width=\columnwidth]{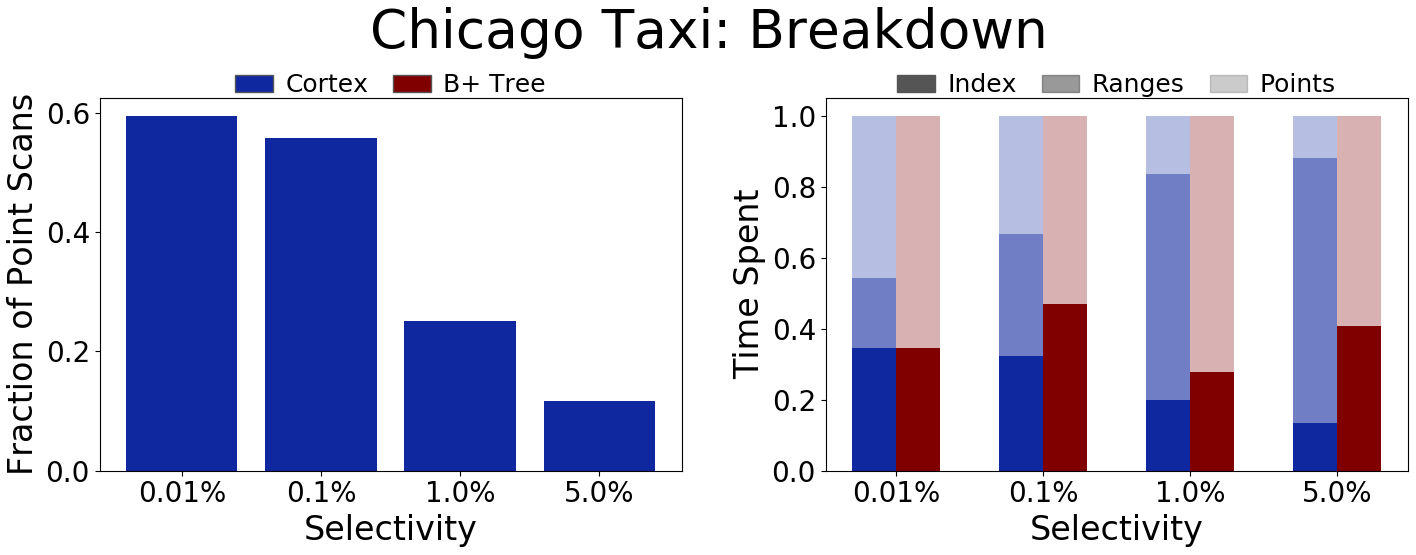}
% Show how just \TheSystem trades out ranges for point scans as selectivity increases. Normalize to 1.
\caption{As the selectivity increases, \TheSystem ($\alpha = 0.2$) performs fewer point accesses using the outlier index, as a fraction of result size (left).  \TheSystem switches from point scans to range scans as the query selectivity increases (right).}
\label{fig:eval:cortex_hybrid}
\end{figure}

\NewPara{Selectivity.}
\Table{eval:punchline:selectivity} shows how the relative performance between \TheSystem and baselines changes as the query selectivity varies from 0.01\% to 5\%. Note that for this and all further experiments, we take $\alpha = 1$, which we find strikes a reasonable balance between good performance at less than 20\% the size of a B-Tree. \TheSystem's outlier assignment remains the same for each query workload; in other words, it is not retrained or reconfigured differently for each set of queries.
We present the results as speedups \emph{relative} to the secondary B-Tree index, since it is de facto standard in databases today. Note that the sizes of the indexes are the same as in \Figure{eval:punchline:btree} and \Figure{eval:punchline:octree}; just the query workload has changed. There are two main takeaways.

First, compared to the other baselines, the B-Tree shines in its handling of queries with low selectivity (at 0.1\% and lower). It outperforms Correlation Maps and Hermit by up to several \emph{orders of magnitude} because it quickly identifies matching records without scanning false positives. However, unlike other approaches, \TheSystem is able to achieve query speeds that are generally competitive with the B-Tree on selective queries, and sometimes even outperform it. \Figure{eval:cortex_hybrid} shows that the explanation lies in \TheSystem's hybrid approach: at low selectivities, \TheSystem gets most of its results from the outlier index, and can therefore avoid scanning unnecessarily large ranges. This is because selective queries cover very few target buckets, and each target bucket only has a small number of host buckets it considers inliers.
Given that a B-Tree's performance on these low selectivity queries is what often justifies its high storage overhead, \TheSystem is an attractive alternative, offering similar performance at a fraction of the storage cost.

Second, \TheSystem consistently outperforms the B-Tree and other alternatives at higher selectivities. Note that it is relatively easy to outperform the B-Tree at selectivites of 5\% or higher; at this selectivity, even a full scan is on par with it. The poor performance of secondary indexes at high selectivities usually results in RDMSes falling back to a full scan for larger query ranges~\cite{secondary-access-path}. However, \TheSystem is able to maintain a $1.5-6\times$ improvement over the B-Tree while also outperforming other indexes. This again follows from \Figure{eval:cortex_hybrid}, since \TheSystem's design lets it transition to efficient range scans and away from costly point scans as the query selectivity grows. This transition is completely organic: even though \TheSystem selects more outliers to scan than on low selectivity queries, a larger fraction of them are subsumed within the inlier ranges being scanned (which are also larger) and are therefore deduplicated. Although deduplication is extra overhead in \TheSystem (as captured by ``Index'' time in \Figure{eval:cortex_hybrid}), at larger selectivities, this cost is more than made up for by avoiding expensive point lookups.

%In fact, \Figure{cortex-high-sel} shows that on the Chicago Taxi dataset, \TheSytem can handle queries up to XXX\% while being more performant than a full scan. This means that \TheSystem is more useful than the B-Tree on a wider range of queries.

Overall, \TheSystem achieves good performance across the board, on both low and high selectivity workloads, compared to both the B-Tree and other baselines.

\NewPara{Varying $\boldsymbol{\alpha}$.} \Figure{eval:punchline:btree} and \Figure{eval:punchline:octree} show that \TheSystem's performance generally improves as $\alpha$ decreases, which is expected, since a lower $\alpha$ means we are willing to give up more storage for even marginally faster performance. As a result, \TheSystem prunes out large ranges of points by assigning as outliers exactly those points that would have otherwise caused it to scan extra host buckets.
\Figure{eval:alpha_points} confirms that as $\alpha$ decreases, the number of point scans (as a fraction of the result size) increases, but is more than compensated by the decrease in the size of the ranges scanned. Indeed, \Figure{eval:alpha_points} shows that stashing a small number of points in the outlier buffer can dramatically reduce the range scans \TheSystem performs.

It is possible that a smaller value of $\alpha$ may actually result in \emph{slower} performance, as seen in \Figure{eval:punchline:btree}. This suggests that \TheSystem's cost model, particularly $\beta$, does not perfectly describe the performance characteristics for each dataset. The $\beta$ we use for this evaluation is an average over multiple datasets, both synthetic (see below) and real, and is customized to each dataset. This suggests further tuning to the data is possible by recomputing $\beta$ for each dataset (and upon inserts); we leave this exploration to future work.

\NewPara{Multi-dimensional Hosts.} The choice of host index plays an important role in the performance of \TheSystem and other baselines. \Figure{eval:punchline:octree} shows that on some datasets, like Stocks, \TheSystem shows only modest performance improvements over CM; on others, like Chicago Taxi and WISE, CM performs well but is still about $4\times$ slower than \TheSystem.
The reason for this lies in the geometry of the host pages. For example, in the Stocks dataset, since the Octree indexes two dimensions, (a) each bucket is wider along the \textit{Open Price} column than on a single-dimensional index and (b) a single value of \textit{Open Price} may map to multiple host buckets. The effect of (a) is to decrease the number of outliers, since nearby values are more likely to be mapped to the same bucket. This is particularly true in the Stocks dataset, where all points lie close to the best fit line. With fewer outliers, CM's performance improves. The effect of (b) is to increase the size of the bucket mapping maintained by the inlier index. The number of host buckets per target bucket increases by $11.1\times, 6.1\times,$ and $12\times$ on the Stocks, Chicago Taxi, and WISE datasets, respectively, when changing from a single to multi-dimensional host index.

Hermit's performance deteriorates substantially when using a multi-dimensional host index because its linear mapping does not understand the division of points imposed by a multi-dimensional index. The outlier tolerance around the piecewise model ends up intersecting a large number of buckets, even at a low selectivity, and even if those buckets do not contain any relevant points. This incurs a larger unnecessary scan overhead than CM, since CM will only scan a host bucket if it is known to contain at least one point in a target bucket touched by the query.

\begin{figure}[t!]
    \centering
    \includegraphics[width=0.8\columnwidth]{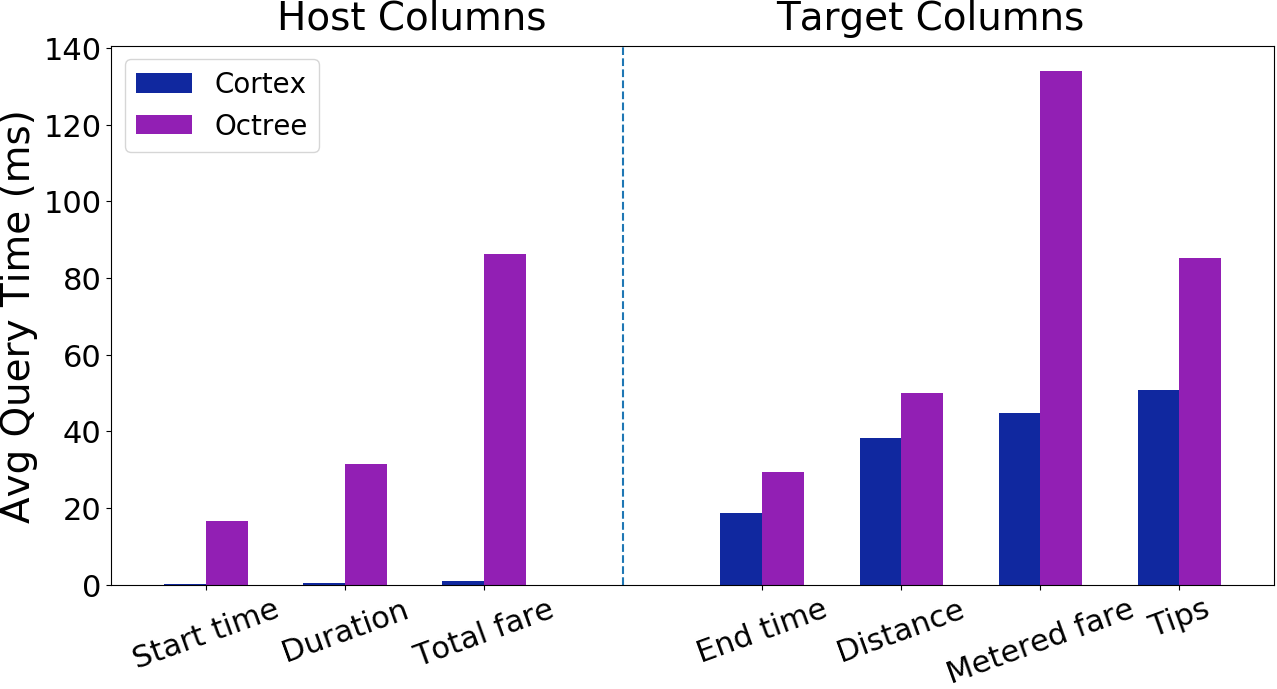}
    \caption{Compared to an Octree on both host and target columns on Chicago Taxi, \TheSystem with an Octree host boosts query times on \textit{all} columns (0.1\% selectivity). Host columns see a $76\times$ boost in query performance on average. For reference, a full scan takes 1.3s.}
    \label{fig:eval:octree_full}
    \vspace{-0.1in}
\end{figure}

Given \TheSystem's performance with single and multi-dimensional host indexes, what is the advantage of a multi-dimensional host index?
Their benefit is in allowing the database to efficiently answer queries along more attributes than if using a single-dimensional host index. With more indexed columns, there's a better chance that a correlation exists that \TheSystem can take advantage of. \Table{eval:datasets} shows that with an Octree host index, we are able to correlate more columns to the set of host columns.

Multi-dimensional indexes also have the option of extending their reach simply by indexing more columns. A natural question is then: how does \TheSystem, with an Octree index over a small number of host columns, compare to using a ``full'' Octree that indexes all columns together (both host and correlated target columns).
\Figure{eval:octree_full} breaks down the performance of these two options on the Chicago Taxi dataset, using queries with 0.1\% selectivity. Though \TheSystem achieves faster query times on the target columns, this is not where its main advantage lies. On the host columns, \TheSystem achieves a $76\times$ speedup over a full Octree. This is a consequence of the curse of dimensionality: by indexing the extra four target columns, the Octree loses granularity on the host columns, making queries on the host columns much slower, consistent with \Figure{corr:octree}. This is particularly undesirable if the host columns happen to contain frequently queried attributes. \Figure{eval:octree_full} thus shows that \TheSystem is effective at extending the reach of host indexes, especially multi-dimensional indexes, to many more attributes.

\NewPara{Correlation Types.}
\Table{eval:punchline:selectivity} suggests that the type of correlation impacts \TheSystem's performance improvement relative to alternatives. \TheSystem (and Hermit) enjoy strong performance on the Stocks dataset (\Figure{eval:punchline:btree}), even on low-selectivity queries, since the strong correlations mean that they can zero in on a narrow range on the host column with few false positives. The Chicago Taxi and WISE datasets have progressively weaker correlations, so \TheSystem's speedup over alternative indexes widens, as those approaches are less effective at indexing weak correlations. Notably, \TheSystem still remains competitive with the B-tree.

% This is a direct result of weaker correlations requiring \TheSystem's inlier index to scan a larger host column range. Indeed, we find that, on a 1-D host index, each target bucket in the Stocks dataset maps to 0.55\% of the total host buckets on average, but that number increases to 2.0\% for the Chicago Taxi dataset, and 11.9\% for WISE.

\begin{figure}[t!]
    \centering
    \includegraphics[width=0.49\columnwidth]{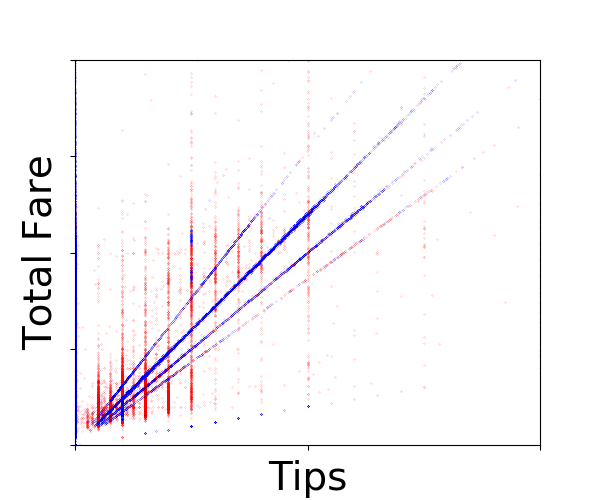}
    \includegraphics[width=0.49\columnwidth]{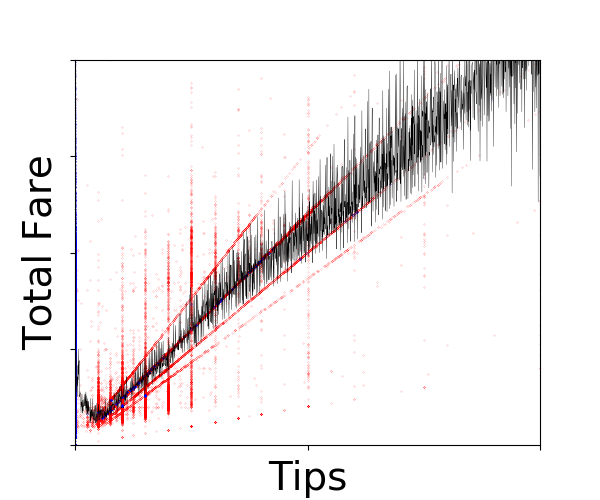}
    \caption{Outliers in \TheSystem (left, $\alpha=0$) and Hermit (right, with the piecewise model), on a correlation in Chicago Taxi.}
    \label{fig:eval:hermit-taxi-fail}
    \vspace{-0.2in}
\end{figure}

In particular, CM deteriorates rapidly in the presence of outliers (consistent with findings in~\cite{hermit}), while Hermit's outlier thresholds are hard-coded for a situation where each target column value maps to only a handful of host column values.
This is particularly noticeable on non-functional correlations, like in the Chicago Taxi dataset shown in  \Figure{eval:hermit-taxi-fail}. Hermit struggles to fit its piecewise linear model to the data. It jumps across large ranges of values, which causes it to scan large ranges of data even on low selectivity queries (confirmed by \Figure{eval:alpha_points}). On the other hand, \TheSystem is able to better isolate the major trend lines. 
\Section{eval:outliers} takes a closer look at the effect of correlation strength on \TheSystem's performance.

In addition to non-functional correlations, \TheSystem is also able to handle correlations between more than two attributes, or \textit{multi-way} correlations.
The WISE dataset offers a straightforward instance of a multi-way correlations between its three coordinate systems: RA / Dec., Galactic Lat/Lon, and Ecliptic Lat/Lon.
Indexing the two coordinates in any one system determines the coordinates in the other two. However, given only one coordinate, the distribution of any other coordinate appears nearly uniformly random. Compared to a host index that only indexes a single coordinate, using a host index with a pair of coordinates results in \TheSystem classifying $4.4\times$ fewer outliers and reducing the number of host buckets per target bucket by $8\times$, when indexing any of the remaining coordinates. In other words, when both coordinates are host columns, \TheSystem can scan much smaller ranges while taking considerably less space. This is how \TheSystem takes advantage of multi-way correlations.

% TODO(vikram): Talk about the WISE data when we have it.

 %\Figure{eval:punchline:corr_type} shows the performance of \TheSystem and baselines on each attribute queried in the Stocks and Chicago taxi workloads, using a 1-D clustered host index. The performance of \TheSystem relative to a secondary index depends largely on the strength of the correlation between the queried column and columns in the host index. For example, since the correlations in the Stocks dataset are strong (\Figure~\ref{eval:stocks:corrs}), \TheSystem is faster a secondary B-Tree index even on very selective queries: the strength of the correlation means \TheSystem (and prior work) can identify a right range on the host column with few false positives. The Chicago Taxi dataset has weaker correlations than the Stocks dataset, but still sufficiently strong for \TheSystem to stay on par with a B-Tree, even on selective queries. On the other hand, the correlations in the WISE dataset are much weaker, so each target bucket maps to more host buckets, which results in a larger range that must be scanned. Indeed, we find that on average, each correlation in the Stocks dataset maps to 0.55\% of the total target buckets, 2.0\% for the Chicago Taxi dataset, and 11.9\% for WISE.

\subsection{\TheSystem's Outlier Assignment}
\label{sec:eval:outliers}

\begin{figure}[t!]
    \centering
    \includegraphics[width=0.49\columnwidth]{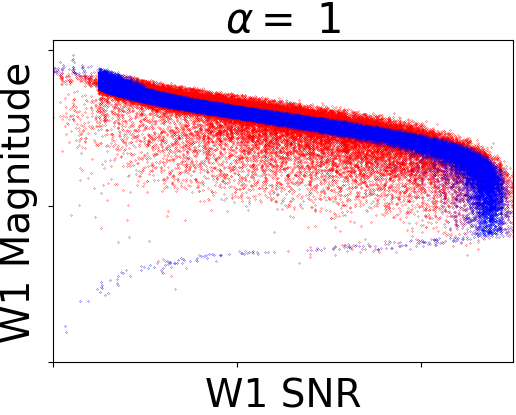}
    \includegraphics[width=0.49\columnwidth]{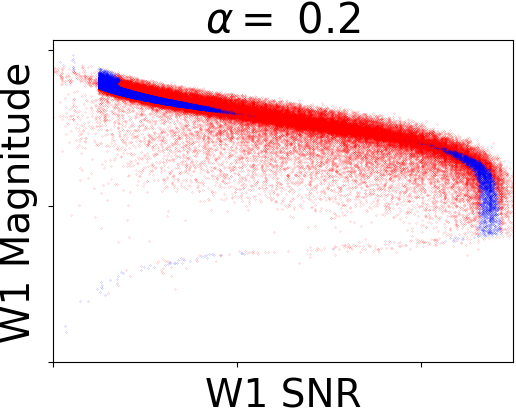}
    \caption{\TheSystem's outlier assignment on a correlation from the WISE dataset. Outliers are shown in red.}
    \label{fig:eval:outliers:examples}
    \vspace{-0.1in}
\end{figure}

Outliers are central to \TheSystem: the choice of outliers almost single-handedly determines the storage-performance tradeoff. In this section, we examine which points \TheSystem prioritizes when making its assignment. \Figure{eval:outliers:examples} helps answer this question by looking at the outliers from a weak correlation in the WISE dataset, between \textit{$W_1$ SNR} and \text{$W_1$ Magnitude}, for two values of $\alpha$. At $\alpha = 1$, \TheSystem eliminates points that do not lie in the main trend of the data, therefore isolating the dense areas. It seems to treat all parts of the trend shape equally: the trend is not substantially wider in some areas than others. However, when decreasing $\alpha$ to 0.2, \TheSystem assigns many more outliers to the center of the trend than to other parts. The reason is that the trend has a shallower slope in the middle. A host bucket (of \textit{$W_1$ magnitude}) is intersected by more target buckets in the center of the trend than at the endpoints. \TheSystem realizes that these host buckets add a large number of false positives and increase scan overhead. In response, \TheSystem tends to stash points in these host buckets into the outlier index. Importantly, this shows that \TheSystem therefore can prioritize different areas of the correlation, and assign proportionally more or fewer outliers as necessary. This distinguishes \TheSystem from prior work: the ability to flexibly assign outliers is only possible because \TheSystem doesn't determine outliers based on a fixed functional model.

% How is \TheSystem still able to outperform them, sometimes by up to $2\times$ (\Table{eval:punchline:selectivity})?
% The key is in how \TheSystem chooses its outliers. \Figure{eval:outliers:stocks_sample} shows that even on a strongly linear correlation from the Stocks dataset, \TheSystem is able to choose outliers in a way that \emph{isolates dense areas} of the data. This means that \TheSystem will clear out as many points $p$ as possible that occupy the same host bucket as these dense regions; this makes sure that a query over $p$'s target bucket does not incur the overhead of scanning the entire dense region. This targeted approach to assigning outliers is only possible because of \TheSystem's model-free approach: instead of modelling the data with a hard-coded function and using a single threshold to classify points as outliers, \TheSystem drops any assumptions about the nature of the correlation. Instead, it only chooses outliers that will maximize its performance.

% \NewPara{Where are the outliers?}
% To qualitatively understand \TheSystem's outlier assignment, \Figure{eval:outliers:examples} presents sample two-way correlations from our real datasets, along with the points that \TheSystem classifies as outliers (in red) for $\alpha = 1$. Points in less dense areas typically tend to be classified as outliers, but this is not strictly true: outliers may be chosen in denser regions of the data if doing to avoids the overhead of scanning a large host bucket.

\subsection{Scalability}
\label{sec:eval:scalability}

In this section, we probe \TheSystem's performance when extended to many columns and a large amount of the noise in the correlation.
We use a synthetic dataset with 100M records with values in $[0, 10^6]$. Correlations consist of an exact linear map $(i.e., y = x)$ with some \textit{noise fraction} $f$ (which varies by experiment). For a given $f$, we add additive noise to $f$ fraction of the points, drawn from a Laplace distribution with $\sigma = 200000$ (one-fifth of the value range). 

\begin{figure}[t!]
    \centering
    \includegraphics[width=0.49\columnwidth]{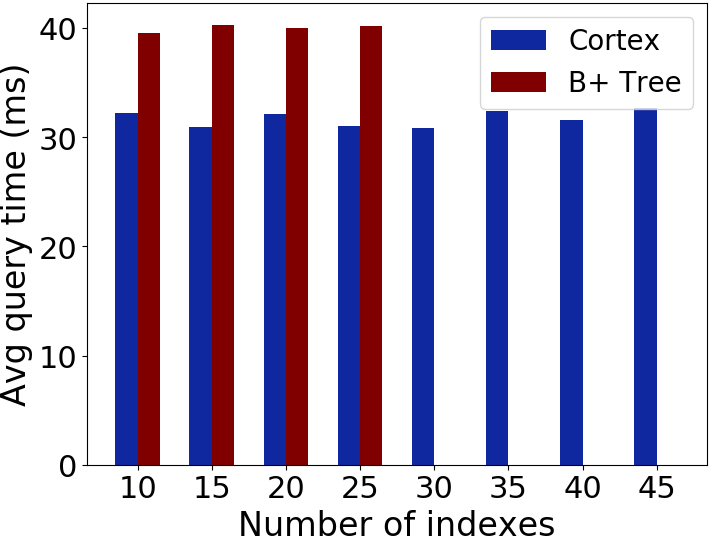}
    \includegraphics[width=0.49\columnwidth]{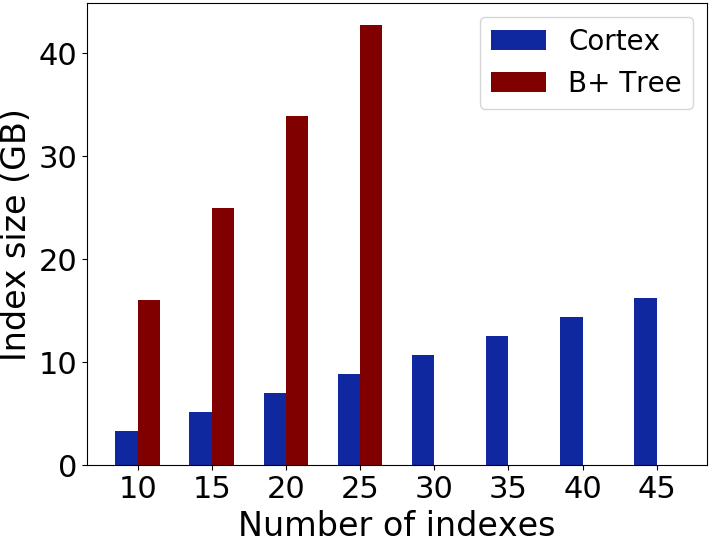}
    \caption{\TheSystem scales to more columns (without sacrificing query speed) than a traditional secondary index, which runs out of memory after indexing 25 columns.}
    \label{fig:eval:scale_columns}
    \vspace{-0.1in}
\end{figure}

\NewPara{Number of Columns.}
\TheSystem scales linearly, both in space usage and performance, with the number of target columns, since each column is indexed independently of the others, much like a secondary index.
\Figure{eval:scale_columns} confirms this expectation using our synthetic dataset. We generate increasingly larger datasets with between 10 and 45 columns, using noise fraction $f = 0.2$. For each dataset, we create \TheSystem ($\alpha=1$) and B-Tree indexes on all except a one column (the host column). Since \TheSystem occupies a fraction of the space of a B-Tree, we found that the B-Tree ran out of memory after indexing 25 columns. Meanwhile, \TheSystem scales to 45 columns.

\begin{figure}[t!]
    \centering
    \textbf{Robustness to Noise} \\
    \includegraphics[width=0.49\columnwidth]{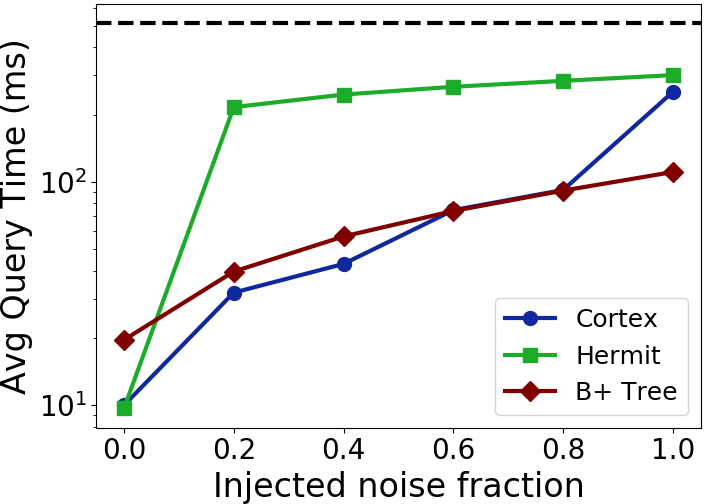}
    \includegraphics[width=0.49\columnwidth]{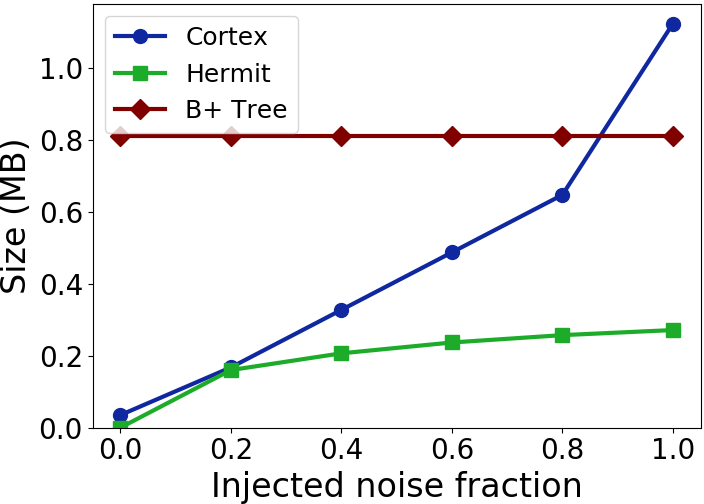}
    \caption{\TheSystem's query performance and speed relative to baselines, as a function of the noise fraction. The dotted line indicates a full scan. Note the log scale on the left.}
    \label{fig:eval:noise_inject}
    \vspace{-0.1in}
\end{figure}

\NewPara{Robustness to Noise.}
How weak of a correlation can \TheSystem index and still maintain a favorable size-speed tradeoff when compared with secondary indexes? To quantify this, we run \TheSystem and a secondary B-Tree index on our synthetic dataset with 2 columns and increasingly more noise, measured by the noise fraction $f$ defined above. \Figure{eval:noise_inject} shows that \TheSystem is able to maintain performance on par with a B-Tree up until the data is 80\% noise, all while occupying a fraction of the B-Tree's memory footprint. As long as a strong correlation exists under the noise, \TheSystem finds it by stashing all other points as outliers. Indeed, \TheSystem's space usage grows linearly with the noise fraction, reflecting its ability to grow its outlier index without tuning. On the other hand, Hermit's performance deteriorates since it does not locate as many outliers. This is a consequence of Hermit's requirement of a hard-coded limit on outliers, which doesn't let its outlier stash adapt to the correlation like \TheSystem's. Note that \TheSystem's deterioration in performance at 100\% noise is due to two factors. \TheSystem only stashes 47\% of the records, as opposed to $f = 0.8$ where it stashed 80\% of records, which means its outlier assignment cannot capture all the noise. Even though the number of outliers dropped, \TheSystem's size still increases; this is because with fewer outliers stashed, each target bucket now maps to many more host buckets, increasing the size of the cell mapping maintained by the inlier index.

 \begin{figure}[t!]
    \centering
    \includegraphics[width=0.49\columnwidth]{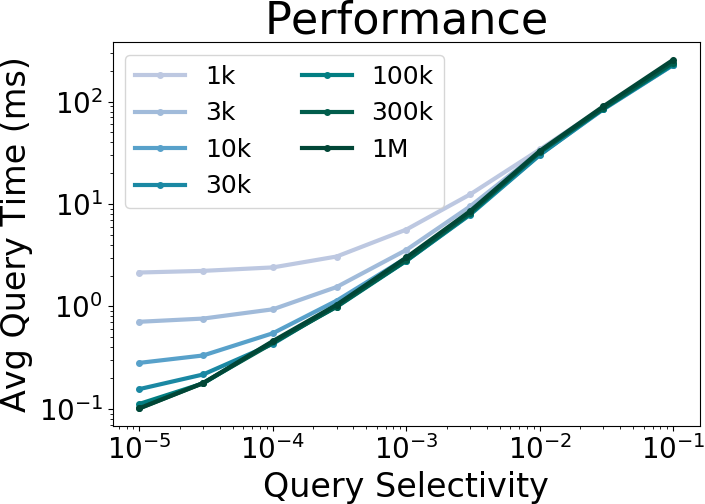}
    \includegraphics[width=0.49\columnwidth]{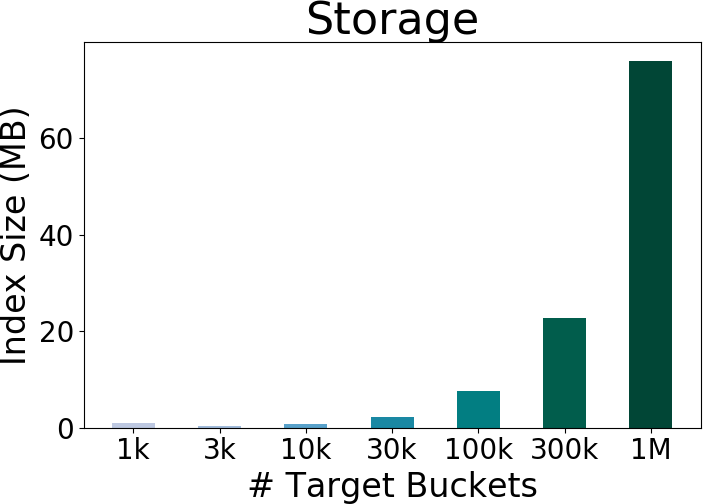}
    \caption{Increasing the number of target buckets lets \TheSystem handle more selective queries but uses more space.}
    \label{fig:eval:target_buckets}
    \vspace{-0.1in}
\end{figure}

\begin{figure}[t!]
    \centering
    \includegraphics[width=0.49\columnwidth]{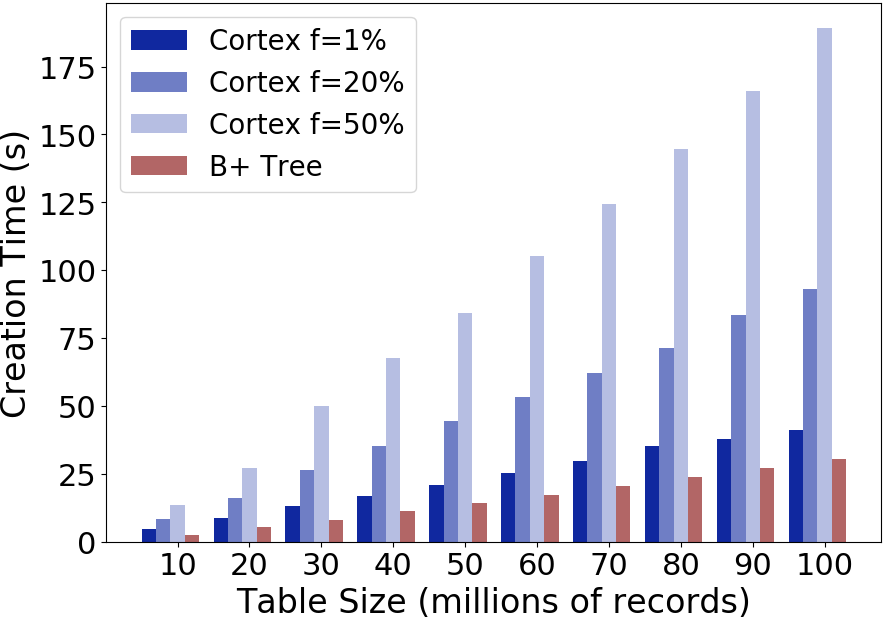}
    \includegraphics[width=0.49\columnwidth]{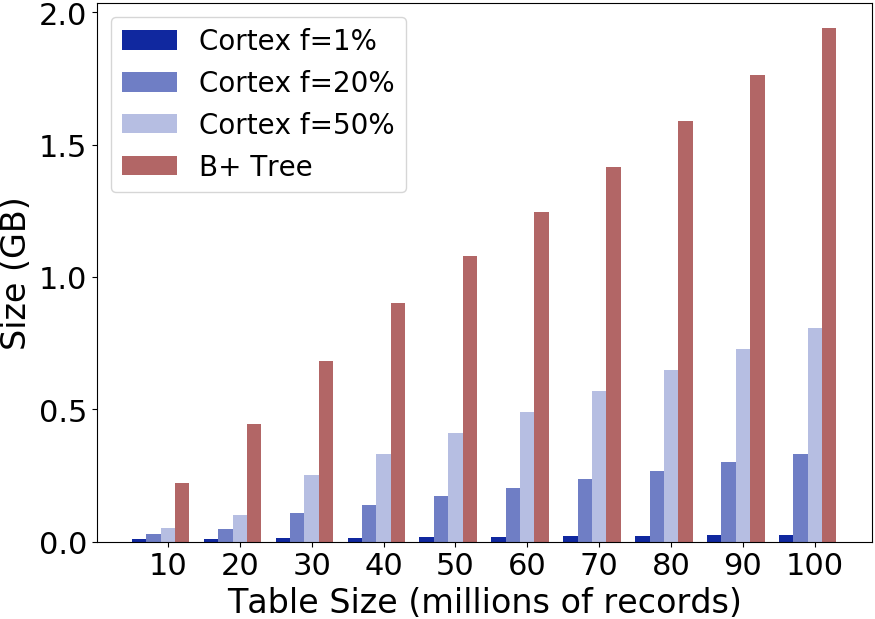}
    \caption{Creation time (left) and size of \TheSystem (right) over a range of table sizes and correlation strengths.}
    \label{fig:eval:creation}
    \vspace{-0.1in}
\end{figure}

\NewPara{Bucketing Strategy.} 
\label{sec:eval:target_buckets}
\Section{algo} outlines \TheSystem's default to use a target bucket size that matches the width of the lowest selectivity queries. Here, we examine the reasoning behind this choice. We vary the target bucket size and observe the query performance over a range of selectivities in \Figure{eval:target_buckets}. We find that queries at selectivity $s$ do not experience a slowdown as long as there are at least $\frac{1}{s}$ buckets. Any fewer, and the target buckets become too wide: a query that only slightly overlaps a target bucket must still incur the overhead of scanning all its inlier host buckets. Importantly, there is little to no cost to performance for choosing more target buckets. However, the storage overhead scales linearly with the number of target buckets. Therefore, we choose the smallest bucket size that yields good performance for the lowest selectivity query in the workload. If the lowest selectivity is $s_L$, \TheSystem creates $\frac{1}{s_L}$ target buckets.

% \begin{figure}
%     \centering
%     \includegraphics{}
%     \caption{Show how as the coalesced buckets increase in range, their total space decreases because of deduplication.}
%     \label{fig:eval:coalesce}
% \end{figure}

%\NewPara{Coalescing Target Buckets.}
%To ameliorate the effect of small target buckets on high selectivity queries, \Section{algo:coalesce} introduces an optimization to coalesce buckets. The cost of deduplication is substantial for high selectivity queries, and coalescing buckets reduces that cost by pre-aggregating groups of target buckets. Most importantly, coalescing can actually result in \emph{fewer} outliers returned by the outlier index, since it front-loads deduplication onto the pre-aggregation step. The number of outliers that can be deduped ahead of time in this manner depends on the correlation. However, \Figure{eval:coalesce} shows that on the synthetic Laplace dataset, \TheSystem's coalesced buckets contain substantially fewer outliers than the sum of the buckets they cover. This, in turn, improves performance for queries that have to scan a large range of target buckets, specifically high-selectivity queries, at the cost of extra storage. \Figure{eval:target_buckets}(b) shows that coalescing can therefore blunt the downside of choosing smaller target buckets.

\subsection{Insertions}
\label{sec:eval:inserts}

\begin{figure}[t!]
    \centering
    \includegraphics[width=0.49\columnwidth]{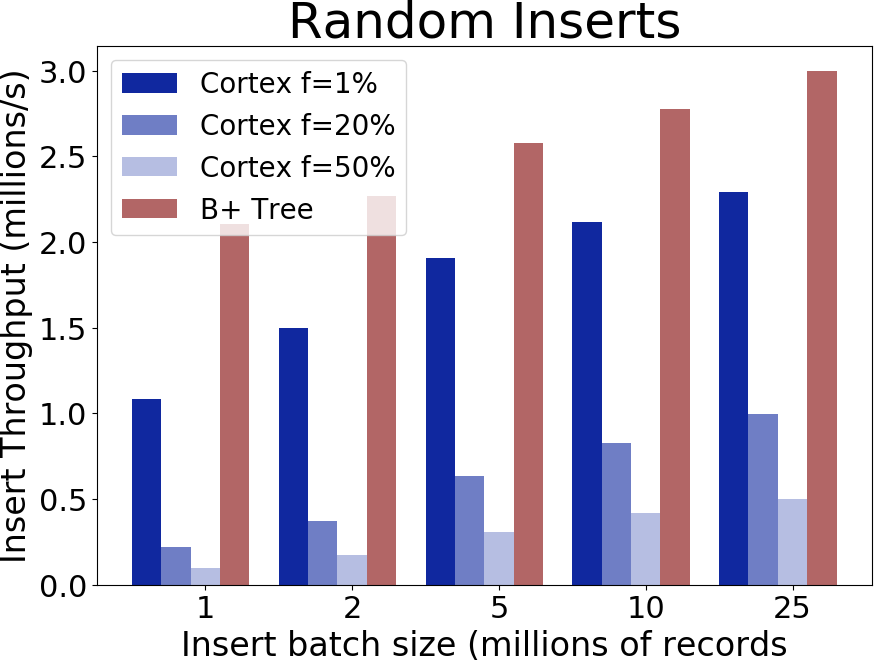}
    \includegraphics[width=0.49\columnwidth]{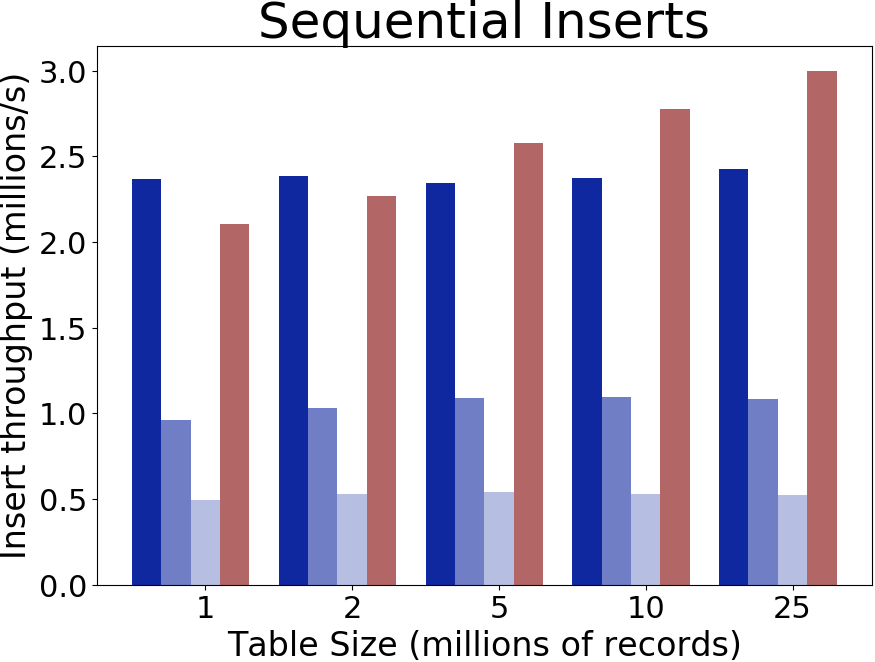}
    \caption{Throughput on random (left) and sequential (right) inserts for a variety of correlation strengths.}
    \label{fig:eval:updates}
    \vspace{-0.2in}
\end{figure}

\TheSystem's creation consists of initializing its correlation tracker with all the cells that contain points. The latency of this operation depends heavily on the strength of the correlation being indexed. \Figure{eval:creation} shows that strong correlations ($f = 1$\% noise) can be initialized more than $4\times$ faster than weaker than weak correlations ($f = 50$\% noise). Note that \TheSystem achieves more than 1M inserts per second on strong correlations, but is slower than inserting into a traditional B-Tree, since it must first aggregate records by cell.

\TheSystem allows incremental updates without the need to reassign all records in the table. Since it is often the case that records are inserted in order of primary key (e.g., a timestamp), we evaluate inserts on both a random load, and a load where records are sequentially added in increasing key order. \Figure{eval:updates} shows that, in general, sequential inserts are faster than random inserts; insertion time scales with the number of affected \textit{cells}, and sequential inserts affect fewer host buckets in each batch. This also explains why random inserts are more efficient in larger batches: as more points are added, many of them (particularly inliers) fall in the same bucket.
\section{Conclusion}
\label{sec:conclusion}

This work presented \TheSystem to harness the correlations between attributes in a dataset. As an alternative to a secondary index, which incurs high storage overhead, \TheSystem adapts itself to the primary index on the table, whether single or multi-dimensional, to encode a correlated column in terms of a set of host columns. On queries with low selectivities, it is able to match the B-Tree's performance using $10\times$ less space, while outperforming prior solutions by close to an order of magnitude. On high selectivity queries, it outperforms both B-Trees and other baselines. Key to \TheSystem's performance gains is its ability to judiciously choose which records to consider as outliers. It makes this determination without the use of any predefined model, instead using a cost model of performance to guide its outlier assignment. \TheSystem offers speedups on both single and multi-dimensional host indexes and is a compelling alternative to secondary indexes in the case of correlated columns.

\bibliographystyle{ACM-Reference-Format}
\bibliography{scorpia_refs}
\end{document}